\documentclass[a4paper, 11pt]{article}
\usepackage{jheppub}
\usepackage{appendix} 
\usepackage{amssymb,amsmath} 
\usepackage{amsthm}
\usepackage{mathtools}
\usepackage{textcomp}
\newcommand{\Tr}{\,\mathrm{Tr\,}}

\title{Studying superconformal symmetry enhancement through indices}
\author{Mikhail Evtikhiev
\\
\it{Department of Particle Physics and Astrophysics,\\
Weizmann Institute of Science, Rehovot 7610001, Israel}
}
\emailAdd{Mikhail.Evtikhiev@weizmann.ac.il}
\abstract{
	In this note we classify the necessary and the sufficient conditions that an index of a superconformal theory in $3\leq d \leq 6$ must obey for the theory to have enhanced supersymmetry. We do that by noting that the index distinguishes a superconformal multiplet contribution to the index only up to a certain equivalence class it lies in. We classify the equivalence classes in $d=4$ and build a correspondence between ${\cal N} = 1$ and ${\cal N}>1$ equivalence classes. Using this correspondence, we find a set of necessary conditions and a sufficient condition on the $d=4$ ${\cal N} = 1$ index for the theory to have ${\cal N}>1$ SUSY. We also find a necessary and sufficient condition on a $d=4$ ${\cal N}>1$ index to correspond to a theory with ${\cal N} > 2$. We then use our results to study some of the $d=4$ theories described by Agarwal, Maruyoshi and Song, and find that the theories in question have only ${\cal N} = 1$ SUSY despite having rational central charges. In $d=3$ we classify the equivalence classes, and build a correspondence between ${\cal N} = 2$ and ${\cal N}>2$ equivalence classes. Using this correspondence, we classify all necessary or sufficient conditions on an $1\leq {\cal N}\leq 3$ superconformal index in $d=3$ to correspond to a theory with higher SUSY, and find a necessary and sufficient condition on an ${\cal N} = 4$ index to correspond to an ${\cal N} > 4$ theory. Finally, in $d=6$ we find a necessary and sufficient condition for an ${\cal N} = 1$ index to correspond to an ${\cal N}=2$ theory. 
}
\begin{document}

\maketitle

\section{Introduction and summary of results}
Superconformal theories have been extensively studied in the past 40 years. This has partly been because their enhanced symmetries allow for various exact computations to be performed in these theories, even at strong coupling, so that they provide useful windows into strong coupling physics. Superconformal field theories exist in $2\leq d\leq 6$ dimensions, and in $3\leq d\leq 6$ dimensions a superconformal algebra is essentially defined by the number of spacetime dimensions $d$ and the amount of supersymmetry ${\cal N}$. Superconformal theories exist for $1\leq {\cal N}\leq 6$, ${\cal N}= 8$ in 3 dimensions, $1\leq {\cal N}\leq 4$ in 4 dimensions, ${\cal N}=1$ in 5 dimensions and ${\cal N}=(1, 0), (2, 0)$ in 6 dimensions.

 In the last 25 years superconformal theories without a known Lagrangian (non-Lagrangian theories) became an important research subject. Interest in the non-Lagrangian superconformal theories was first kindled by the seminal paper \cite{Argyres} and since then this topic was vastly developed. In the absence of a Lagrangian one needs some quantity to describe the theory, and one of the most convenient options is the superconformal index. The superconformal index was first defined by R\"{o}melsberger and Kinney et al. \cite{Romelsberger, Kinney} for ${\cal N} = 1$ $d=4$ theories and since then became a popular tool to study the content of superconformal theories in various spacetime dimensions (see \cite{Bhattacharya, Dolan2} for index definitions in different dimensions, \cite{Romelsberger2, Dolan, Dolan3, Benini, Ardehali, Aharony2, Dimofte, Aharony3, Bashkirov2, Tanzini} for examples of index applications, and \cite{Rastelli} for 4-dimensional index review). Unfortunately, the index captures the short multiplets only up to recombination, so it is impossible to restore the full short multiplet content from the index; one can determine only the net difference between the number of multiplets that recombine together into a long multiplet. Therefore, one can put all the multiplets into certain equivalence classes according to their contribution to the index; this idea was first suggested in \cite{Gadde} for the $d=4$ ${\cal N}=2$ theories and then developed for the ${\cal N} = 1$ case in \cite{Beem}.

The superconformal index might not fully reflect properties of the theory; for example in \cite{Song} many ${\cal N} = 2$ Argyres-Douglas theories were found at the end of an RG flow, triggered by a $d=4$ ${\cal N} = 1$ massive deformation. If one would compute the index in this case, one would get an ${\cal N} = 1$ index for a theory that in fact has ${\cal N} = 2$ SUSY. In this note we study how one can find SUSY enhancement through the index by studying properties of the superconformal index in 3 to 6 dimensions. In other words, we want to find necessary and/or sufficient conditions on a low-SUSY index to correspond to a theory with a higher amount of SUSY. It is reasonable to expect that certain restrictions on the index exist, because if SUSY enhancement happens, the original multiplets should somehow combine to the new, enhanced ones.

There are various ways in which one could enhance supersymmetry, and we focus on two of them. First, we can couple a theory to an exactly marginal deformation (EMD), such that SUSY can be enhanced for a specific value of this deformation. Another possibility is to trigger an RG flow by a massive deformation. 

In 3 dimensions only ${\cal N} = $ 1, 2 admit EMDs, and in ${\cal N}=2$ theories a deformation can be exactly marginal if and only if it doesn't break any flavor symmetry \cite{Dumitrescu, Green}. On the other hand, all ${\cal N}>2$ theories have an ${\cal N}=2$-preserving massive deformation; for ${\cal N}=3,4$ theories it lies in the flavor current multiplet, while for ${\cal N}>4$ theories it lies in the stress-tensor multiplet. Thus, ${\cal N} = 2$ theories can be enhanced to ${\cal N} > 2$ ones by an EMD; one can also possibly obtain an enhanced SUSY by triggering an ${\cal N}=2$-preserving RG flow.

In ${\cal N} = $ 1, 2 four-dimensional superconformal theories the number of EMDs is not fixed, pure ${\cal N} = 3$ theories do not admit EMD preserving the full supersymmetry and in ${\cal N} = 4$ there is always one and only one EMD that preserves the full SUSY. Thus, some theories may have their SUSY enhanced by EMD; in particular, ${\cal N} = 1$ can be enhanced to $2\leq {\cal N}\leq 4$, a famous example involves the EMDs that break ${\cal N} = 4$ to ${\cal N} = 1$ \cite{Parkes}. The index does not change under an EMD, so if some theory can be deformed to a theory with higher amount of SUSY by an EMD, the index can't obey any sufficient conditions for enhancement. This also means there cannot exist a necessary and sufficient condition on an ${\cal N} = 1$ index that would show us ${\cal N} = 1 \rightarrow {\cal N} = 2,3,4$ SUSY enhancement. Extra constraints on the index, when part of the enhanced SUSY is visible in the UV, are discussed in appendix \ref{flavN2}.

In 5 dimensions there is only one superconformal algebra and SUSY enhancement is impossible. In 6 dimensions one cannot have relevant or marginal deformations \cite{Dumitrescu, Louis} that would preserve superconformal theory; however we still list a necessary and sufficient condition on SUSY enhancement through index, as one can obtain ${\cal N} = (1,0)$ index for a theory that actually has ${\cal N} = (2, 0)$ SUSY from e.g. stringy construction.

The index as a function does not depend on the RG flow. This means that one can compute it in the UV, then trigger an RG flow and track it to the IR where the superconformal symmetry is hopefully restored. However, the R-charge is not guaranteed to remain the same along the flow, so the function one gets in the IR is not necessarily the IR index and one must do a-maximization \cite{Intriligator} to find the real index in the IR\footnote{This can only be done assuming that the IR R-symmetry is visible in the UV or involves some fields becoming free, and it is not accidental \cite{Kutasov}.}. On the other hand, unlike in the EMD case the index in the IR may obey sufficient conditions for SUSY enhancement, and this was the case e.g. for all the theories studied by Maruyoshi and Song in \cite{Song} when they were able to compute the index (we should notice that the SUSY enhancement conditions suggested in \cite{Song} are not precise, as we will discuss below). 

This note is composed as follows. In section \ref{enh} we list the SUSY enhancement conditions on the $d=4$ index. We show that there is a necessary and sufficient condition for SUSY enhancement on ${\cal N}>1$ indices. We also show there is one sufficient and a set of necessary conditions on the ${\cal N} = 1$ index for SUSY enhancement, and we list them for ${\cal N} = 1 \rightarrow {\cal N} = 2, 3, 4$. Section \ref{ams} is dedicated to using some of the results we obtained in the previous section. We first describe the massive deformation method, following \cite{Song}, and then study several of the theories obtained in \cite{Agarwal}, using the methods we developed for $d=4$ indices, to show how different necessary and sufficient conditions on ${\cal N} = 1 \rightarrow {\cal N} = 2$ enhancement work out in different cases; we prove that several theories found by \cite{Agarwal} have only ${\cal N} = 1$ SUSY despite having rational central charges. 

In section \ref{d3} we list the SUSY enhancement conditions on the $d=3$ index. We show there is a necessary and sufficient condition for SUSY enhancement on ${\cal N}>4$ indices. We demonstrate there is a sufficient and a set of necessary conditions for SUSY enhancement on ${\cal N}=3$ index, we list them for ${\cal N} = 3 \rightarrow 4\leq  {\cal N} \leq 8$. We show there is a set of necessary conditions for SUSY enhancement on ${\cal N}=2$ index and list them for ${\cal N} = 2 \rightarrow 3\leq  {\cal N} \leq 8$. Finally, for ${\cal N} = 1$ index we show there is one necessary condition on SUSY enhancement and describe it.

In section \ref{d6} we formulate a necessary and sufficient condition on ${\cal N} = 1 \rightarrow {\cal N} = 2$ SUSY enhancement in $d=6$. 

Appendices contain derivations of the enhancement conditions and related computations. In appendix \ref{d4comp} we present various derivations for the $d=4$ case. In subsection \ref{multred} we set the notation and then list how ${\cal N} = 2$ multiplets decompose into ${\cal N} = 1$ multiplets under SUSY reduction, in subsection \ref{alg} we describe how the relation between equivalence classes in ${\cal N} = 1$ and ${\cal N}>1$ may be built, and in subsection \ref{algN2} we work out this relation between ${\cal N} = 1$ and ${\cal N} = 2$ equivalence classes in details. In subsection \ref{flavN2} we work out the relation between ${\cal N} = 1$ and ${\cal N} = 2$ equivalence classes with flavor taken into account.

In appendix \ref{d3comp} derivations for the $d=3$ case are presented. In subsection \ref{notapp} we set the notation, in subsection \ref{d3n2} we describe multiplets content in ${\cal N} = 2$, and define ${\cal N}=2$ index. In subsection \ref{n3mult} we shortly describe multiplet content in ${\cal N} > 2$ algebras. In subsection \ref{n2cond} we derive conditions on ${\cal N}=2$ index for SUSY enhancement and in subsection \ref{n3cond} we do the same for ${\cal N}=3$ index.

Throughout the paper we implicitly assume that the theory in question has only one sector (and only one stress tensor)\footnote{We are grateful to Shlomo Razamat for pointing us towards this issue.}. This assumption is generally reasonable. For example, for the case of relevant deformations, when we consider SUSY enhancement in the IR, we can only study theories that do not feature decoupling into different sectors, as otherwise there is also an extra R symmetry in the IR that is not visible in the UV, and we do not know how to compute the IR index. In case of EMDs one can indeed imagine a decoupling into two or more theories for a specific value of the EMD. Unfortunately, in such a case our analysis is not applicable. This is because in the many sectors case only some of the sectors may experience SUSY enhancement, so we can't assume that all lower-SUSY multiplets recombine into the higher-SUSY ones, and therefore necessary conditions on SUSY enhancement we have derived become void, and the sufficient conditions can't be satisfied in the EMD enhancement scenario, as we have argued above. 

In the most general many-sector scenario the sufficient SUSY enhancement conditions still will be valid. However, without additional assumptions (or ability to factorize the index into a product of indices corresponding to the different sectors) one at most can merely find the amount of newly conserved supercurrents, but not which sectors have experienced SUSY enhancement and what is the resulting amount of supersymmetry in each of the sectors.

\section{Index and SUSY enhancement conditions in 4 dimensions} \label{enh}
All conditions on SUSY enhancement from the index can be divided into sufficient, necessary and sanity checks. A sufficient condition shows that the index at hand in fact must correspond to a theory with a higher amount of SUSY (or a free theory). Such a condition can only be related to the presence of an extra supercurrent multiplet. Necessary conditions stem from the requirement on lower-SUSY multiplets to be able to recombine into higher-SUSY multiplets in the case of SUSY enhancement. Finally, sanity checks are employed to verify that the function we study is in fact a superconformal index; they check that index contributions do not violate unitarity bounds.

Obviously, the most desired SUSY enhancement condition would be both necessary and sufficient. For such a condition to exist, we need the extra supercurrent multiplet to be protected. It turns out, that this is the case for the ${\cal N} = 2$ superconformal algebra; a spin-$(\frac12,1)$ supercurrent multiplet called $\mathcal{D}_{\frac{1}{2},(0,0)}$ is protected \cite{Osborn} (our notation for ${\cal N} = 2, 4$ multiplets is in agreement with \cite{Osborn}) and therefore one can write a necessary and sufficient condition on an ${\cal N} = 2,3$ index to correspond to a higher-SUSY theory. Let us define the ${\cal N} = 2$ index as 
\begin{equation}
	\mathcal{I}_2 (t, x, v) = \Tr (-1)^F t^{2+2j+\frac{4r+R_2}{3}} x^{2\bar{j}} v^{R_2/3-2r/3}
\end{equation}
where $r, R_2$ are Cartans of $SU(2)_r \times U(1)_{R_2}$ R-symmetry (normalized in such a way that the supercharge $Q^1$ has $SU(2)_r$ charge $\frac12$ and $U(1)_{R_2}$ charge 1), $j, \bar{j}$ are the charges under $SU(2)_L \times SU(2)_R$ and the trace is taken over all operators. In this case the coefficient $\alpha$ of the term $$\alpha t^{5/3}v^{-1/3} \left(x+\frac1x\right)$$ (that is related to $\mathcal{D}_{\frac{1}{2},(0,0)}$) must be nonnegative. If $\alpha>0$, then there is ${\cal N} = 2 \rightarrow {\cal N} = 2+\alpha$ SUSY enhancement or the theory is free; the converse is also true.

The ${\cal N} = 1$ superconformal algebra has very few protected multiplets (and it doesn't protect conserved current or conserved supercurrent multiplets), so we cannot have a necessary and sufficient condition, and must carry out a more thorough analysis, to which we dedicate the rest of the section. The ${\cal N} = 1$ superconformal index in 4 dimensions was first defined by R\"{o}melsberger; we will use the definition similar to \cite{Dolan}: 
\begin{equation}
	\tilde{\mathcal{I}}(t, x) = \mathrm{Tr}\; [(-1)^F t^{\mathcal{R}}x^{2\bar{J}_3}]
\end{equation}
where only the operators in the kernel of $\mathcal{H}$ contribute, $\mathcal{H} = \frac{1}{2} \{Q^\dagger, Q\}$, $Q$ is a specific supercharge with $U(1)_R$ charge 1 and $SU(2)_L$ charge~$-\frac{1}{2}$, $J_3$ is the spin under $SU(2)_L$, $\bar{J}_3$ is the spin under $SU(2)_R$ part of $SO(4)$, and $\mathcal{R} = R + 2J_3 $ with $R$ being the generator for $U(1)_R$ R-symmetry. For example, according to our definition the contribution of the ${\cal N} = 1$ stress-tensor multiplet is given by 
\begin{equation}
	\frac{-t^3 \left(x+\frac1x\right)}{(1-tx)(1-t/x)}.
\end{equation} 
Let us notice that all ${\cal N} = 1$ single-multiplet contributions look like 
\begin{equation}
	\tilde{\mathcal{I}}_1 = \sum\limits_i \frac{t^{\alpha(i)}\chi_{j(i)} (x)}{(1-t/x)(1-tx)},
\end{equation}
where  $\chi_j$ is an $SU(2)$ spin-$j$ character; for almost all multiplets the sum contains only one term (the exception being free-field multiplets). One can notice that if one multiplies the index by $(1-t/x)(1-tx)$, one gets an infinite series in $t$, that loosely speaking is an infinite series in the scaling dimensions of the ${\cal N} = 1$ multiplets' bottom components. Therefore it is easier to deal with the corrected index 
\begin{equation}
	\mathcal{I} = (1-t/x)(1-tx) (\tilde{\mathcal{I}}-1)
\end{equation}
and search for certain powers of $t, x$ (we subtract 1 because there is no Verma module related to the trivial representation of the superconformal symmetry). We will consider the corrected index from now on, omitting the word ``corrected'' for brevity. In the subsections below we list the results without derivations, some details can be found in appendices \ref{alg}, \ref{algN2}; our notation for ${\cal N} = 1$ multiplets is defined in appendix \ref{multred} and for ${\cal N} = 2,4$ it is consistent with \cite{Osborn}. 

\subsection{SUSY enhancement and restrictions on ${\cal N}=1$ indices for enhancement to ${\cal N}=2$} \label{req}
There is one sufficient and an infinite set of necessary conditions on index for $\mathcal{N}=1$ index to correspond to $\mathcal{N}=2$ theory. We prove in \ref{alg}, \ref{algN2} that due to relation between $\mathcal{N}=1$, $\mathcal{N}=2$ index equivalence classes no other conditions on SUSY enhancement can be derived from the index without making some assumptions about the theory. \\
{\bf Sufficient condition:} If the coefficient in front of the 
\begin{equation}
	t^\frac{7}{3}(x+\frac{1}{x})
\end{equation} term is positive and equals $k$, we have either SUSY enhanced to at least ${\cal N} = k+1$ or a free theory; the converse is not necessarily true. This happens, because the $[\frac{1}{3}, \frac{1}{2}]_+$ equivalence class\footnote{$[a, b]_\pm$ are equivalence classes as defined in \cite{Beem}, where $[a, b]_+$ and $[a, b]_-$ contribute to the same terms in the index with an opposite sign. Detailed discussion can also be found in appendix \ref{algN2}.} contains only the $\hat{\mathcal{H}}_{(0,\frac{1}{2})}$ multiplet (with a conserved spin-$\frac{3}{2}$ current), while the $[\frac{1}{3}, \frac{1}{2}]_-$ class, that contributes with an opposite sign, is not empty.
\paragraph*{Necessary conditions.}
\begin{itemize}
	\item[1. ] If we know that the ${\cal N} = 1$ theory in question possesses a global symmetry $F$ with dimension $\mathrm{dim}(F)$ and $k$ $\bar{\hat{\mathcal{S}}}_0$,\footnote{Notation for $\mathcal{N}=1$ superconformal multiplets can be found in appendix \ref{multred}} matter multiplets, then either the coefficient in front of $t^{\frac43}$ is at least $\mathrm{dim}(F)-3-k$ or there is no ${\cal N} = 1 \rightarrow {\cal N} = 2$ SUSY enhancement (there might be ${\cal N} = 1 \rightarrow {\cal N} = 3, 4$ enhancement). This is due to the fact that in an ${\cal N} = 2$ theory the flavor current lies in the multiplet with the moment map operator that contributes to the ${\cal N} = 1$ index as $t^{\frac43}$, and the only entity that contributes to the index as $-t^{\frac43}$ is the left free chiral multiplet. As at most 3 ${\cal N} = 1$ flavor currents can become ${\cal N} = 2$ R-currents and lie in the ${\cal N} = 2$ stress-tensor multiplet (in addition to the ${\cal N} = 1$ R-current), if the $t^{\frac43}$ coefficient is less than $\mathrm{dim}(F)-3-k$, we have not enough moment map operators and can't have SUSY enhancement. As contributions to $t^{\frac43}$ term also come from ${\cal N} = 2$ EMDs and ${\cal N} = 2$ conserved supercurrent, it is not impossible to have the $t^{\frac43}$ coefficient bigger than $\mathrm{dim}(F)$.
	\item[2. ] If there is a term in the index 
	\begin{align}
		\alpha (-1)^{2j} t^{2+\tilde{R}} \chi_j(x); \qquad \frac{2j-2}{3} > \tilde{R} \geq \frac{2j-4}{3}
	\end{align} 
	then each of the coefficients in front of 
	\begin{equation}
		(-1)^{2j+1} t^{\frac{7}{3}+\tilde{R}} \chi_{j\pm\frac{1}{2}}(x), \qquad (-1)^{2j} t^{\frac{8}{3}+\tilde{R}} \chi_j(x)
	\end{equation} 
	minus the sum of the coefficients in front of terms 
	\begin{equation}
		(-1)^{2j+1} t^{2-\frac{2(j-k)}{3}+\tilde{R}} \chi_{k}(x), \qquad 0 < k \leq j-\frac{1}{2}
	\end{equation} should be greater than $\alpha$, or ${\cal N} = 1 \rightarrow {\cal N} = 2$ SUSY enhancement cannot happen. This is because of all ${\cal N} = 2$ multiplets, only $\mathcal{E}$ can contribute to ${\cal N} = 1$ index these terms, and the expression we described above counts the number of $\mathcal{E}_{\frac{3\tilde{R}+6}{2}, (0, \bar{j})}$. 
	This condition is used most often for $\bar{j}=0$ case, for which it can be simply reformulated as ``If in the index there is a term 
	\begin{equation}
		n t^R; \qquad \frac23<R<\frac43,
	\end{equation}
	then the coefficients in front of the 
	\begin{equation}
		-t^{R+1/3}\left(x+\frac1x\right), \qquad t^{R+2/3}
	\end{equation}
	terms should not be smaller than $n$''.
	\item[3. ] If there is a term in the index 
	\begin{equation}
		\alpha (-1)^{2j} t^{2+\tilde{R}} \chi_j(x), \qquad \frac{2j}{3} > \tilde{R} \geq \frac{2j-2}{3}
	\end{equation} then the coefficient in front of 
	\begin{equation}
		(-1)^{2j+1} t^{\frac{7}{3}+\tilde{R}} \chi_{j+\frac{1}{2}}(x)
	\end{equation} 
	minus the sum of the coefficients in front of terms 
	\begin{equation}
		(-1)^{2j+1} t^{2-\frac{2(j-k)}{3}+\tilde{R}} \chi_{k}(x); \qquad 0 < k \leq j-\frac{1}{2}
	\end{equation} 
	and the number of $\mathcal{E}_{\frac{3\tilde{R}}{2}, (0, \bar{j})}$, $\mathcal{E}_{\frac{3\tilde{R}+3}{2}, (0, \bar{j}+\frac{1}{2})}$ (counted as above), should be greater than $\alpha$. This is because of all ${\cal N} = 2$ multiplets only $\mathcal{E}, \mathcal{B}$ can contribute these terms to the ${\cal N} = 1$ index.
\end{itemize}

\paragraph*{Free theory and sanity checks.}
\begin{itemize}
	\item[1. ] If there is a term in the index 
	\begin{equation}
		t^{2+\tilde{R}} \chi_j(x); \qquad \tilde{R} < \frac{-4+2j}{3},
	\end{equation}
	or a term 
	\begin{equation}
		\alpha (-1)^{2j} t^{2+\tilde{R}} \chi_j(x); \qquad \frac{-4+2j}{3} \leq \tilde{R} < \frac{2j}{3}, \quad \alpha < 0
	\end{equation}
	the index does not correspond to a unitary ${\cal N} = 1$ theory.	
	\item[2. ] If the coefficient in front of 
	\begin{equation}
		(-1)^{2j+1} t^{2+\frac{2j}{3}}\cdot \chi_j (x); \qquad j\geq 1
	\end{equation}
	is positive, we have a free theory; the converse is not true. This happens, because the $[\frac{2j}{3}, j]_+$ equivalence class contains only the $\hat{\mathcal{H}}_{(0,j)}$ multiplet (with a conserved spin-$(j+1)$ current), while $[\frac{2j}{3}, j]_-$ is not empty.
\end{itemize}
It is worth noticing, that in the scenario of ${\cal N} = 1$ SUSY getting enhanced to ${\cal N} = 2$ after an exactly marginal deformation the sufficient condition cannot be satisfied. This is because the EMD does not change the index and if the index manifestly shows the presence of a spin-$\frac{3}{2}$ conserved current for ${\cal N} = 2$ theory, then it should also be present for the original ${\cal N} = 1$ theory, which signals a contradiction. However, we can use the sufficient condition to check the SUSY enhancement for the theories that are the endpoints of an RG flow, as during the RG flow the R-charge can change. Similar reasoning also precludes us from having any kind of sufficient conditions for ${\cal N} = 1 \rightarrow {\cal N} = 3, 4$ SUSY enhancement after EMD.\\
{\bf Conditions with additional assumptions:}\\
If one makes the additional assumption (consistent with all known examples) that ${\cal N}=2$ theories do not have "spinning Coulomb branch operators" (namely $\mathcal{E}_{r,(0,j)}$ multiplets with $j>0$; see \cite{Papageorgakis} for a thorough discussion), then one can obtain an additional necessary condition for SUSY enhancement. It turns out, that if there is a term in the index 
\begin{equation}
	t^{2+\tilde{R}} \chi_j(x); \qquad \frac{2j-2}{3} > \tilde{R} > \frac{2j-4}{3}, \quad j\geq 1,
\end{equation}
then there is no SUSY enhancement. Moreover, coefficients in front of the terms 
\begin{equation}
	t^R, \quad -t^{R+1/2}\left(x+\frac1x\right); \qquad \frac43 > R > \frac23
\end{equation}
should be equal. Using this fact, one can also simplify necessary condition 3.

\subsection{SUSY enhancement and restrictions on ${\cal N}=1$ indices for enhancement to ${\cal N}=3$} \label{3req}
Similarly to the previous case, there is one sufficient and an infinite set of necessary conditions on index for $\mathcal{N}=1$ index to correspond to $\mathcal{N}=3$ theory, and due to $\mathcal{N}=1 \leftrightarrow \mathcal{N}=3$ equivalence classes matching (see appendix \ref{alg}) no other conditions on SUSY enhancement can be derived from the index without making some assumptions about the theory. As the obtained results are largely similar to the previous case, we list only those that are new for ${\cal N} = 3$. One can also take in consideration that for a theory to have ${\cal N} = 3,4$ SUSY the central charges $a$ and $c$ must match \cite{Aharony}.\\
{\bf Sufficient condition.} It is same as above.
\paragraph*{Necessary conditions.}
\begin{itemize}
	\item[1. ] If the coefficient in front of the $t^2$ term is negative and less than $-8$, we cannot have SUSY enhanced to ${\cal N} = 3$. This is because this term gets contributions from $\mathcal{S}_2$ multiplets (with a positive sign) and $\hat{\mathcal{H}}_{(0,0)}$ (with a negative sign). The latter hosts ${\cal N} = 1$ conserved flavor currents; as ${\cal N} = 3$ theories do not admit conserved flavor currents \cite{Aharony}, the ${\cal N} = 1$ index of an ${\cal N} = 3$ theory can have at most $9-1=8$ conserved flavor currents, that are in fact ${\cal N} = 3$ R-currents.
	\item[2. ] If the coefficient in front of the $t^{4/3}$ term is less than 2, we cannot have a pure ${\cal N} = 3$ theory. This is because this term gets contributions from $\mathcal{S}_{4/3}$ multiplets (and only from them) with a positive sign, and the ${\cal N} = 3$ stress-tensor multiplet has 2 such multiplets in its ${\cal N} = 1$ decomposition. It can be greater than 2 due to contributions of $\mathcal{B}_{[0; 0], R_3 = 12, (0, 0)}$,\footnote{an ${\cal N} = 3$ multiplet that is canceled under the action of all $\bar{Q}$; in terms of ${\cal N} = 2$ multiplets $$\mathcal{B}_{[0; 0], R_3, (0, \bar{j})} \rightarrow \mathcal{E}_{\frac{R_3}{6}, (0, \bar{j})} \oplus \mathcal{E}_{\frac{R_3}{6}+\frac12, (0, \bar{j}\pm \frac12)} \oplus \mathcal{E}_{\frac{R_3}{6}+1, (0, \bar{j})}$$} multiplets. It could be theoretically less than 2 due to contributions of free field multiplets, but a pure ${\cal N} = 3$ theory cannot have free fields.
	\item[3. ] If the coefficient in front of $t^{4/3}$ is $\alpha>2$, then the coefficient in front of 
	\begin{equation}
	 	-t^{5/3}(x+1/x)
	\end{equation}
	should be at least $2\alpha - 2$, or we cannot have an ${\cal N} = 3$ theory. This is because this term gets contributions from $\mathcal{S}_{5/3 (0, \frac12)}$ multiplets (and only from them) with a negative sign, and the only multiplets allowed in an ${\cal N} = 3$ theory that have $\mathcal{S}_{5/3 (0, \frac12)}$ in their decomposition are the ${\cal N} = 3$ stress-tensor multiplet, $\mathcal{B}_{[0; 0], 12, (0, 0)}$ and $\mathcal{B}_{[0; 0], 18, (0, \frac{1}{2})}$. The stress-tensor has 2 $\mathcal{S}_{5/3 (0, \frac12)}$ and 2 $\mathcal{S}_{4/3}$ in its ${\cal N} = 1$ decomposition, $\mathcal{B}_{[0; 0], 12, (0, 0)}$ has 2 $\mathcal{S}_{5/3 (0, \frac12)}$ and 1 $\mathcal{S}_{4/3}$, while $\mathcal{B}_{[0; 0], 18, (0, \frac{1}{2})}$ has only 1 $\mathcal{S}_{5/3 (0, \frac12)}$. 
	\item[4. ] If there is a term $\alpha t^\gamma$ in the index with $\frac{2}{3}<\gamma<\frac{4}{3}$, then there also should be terms 
	\begin{align}
		&\alpha_1 t^{\gamma+1/3} (x+1/x), \; \alpha_2 t^{\gamma+2/3}, \; \alpha_2' t^{\gamma+2/3} (x^2+1+1/x^2), \; \alpha_3 t^{\gamma+1} (x+1/x); \notag \\
		 &-\alpha_1 \geq 2\alpha, \; -\alpha_3 \geq 2\alpha, \;\alpha_2 \geq 3\alpha, \; \alpha_2' \geq \alpha; \qquad \alpha > 0.
	\end{align}
	 This is because $t^\gamma$ for $\frac{2}{3}<\gamma<\frac{4}{3}$ corresponds to an $\mathcal{S}_\gamma$ multiplet, that only can lie in an ${\cal N} = 3$ $\mathcal{B}_{[0; 0], 9\gamma, (0, 0)}$ multiplet. $\mathcal{B}_{[0; 0], 9\gamma, (0, 0)}$ decomposes into ${\cal N} = 1$ multiplets as 
	 \begin{equation}
	 	\mathcal{B}_{[0; 0], 9\gamma, (0, 0)} \rightarrow \mathcal{S}_\gamma \oplus 2 \mathcal{S}_{\gamma+1/3, (0,\frac12)} \oplus 3 \mathcal{S}_{\gamma+2/3} \oplus \mathcal{S}_{\gamma+2/3, (0,1)} \oplus 2 \mathcal{S}_{\gamma+1, (0,\frac12)} \oplus \mathcal{S}_{\gamma+4/3}.
	 \end{equation}

\end{itemize}
Similarly to section \ref{req}, in the scenario of ${\cal N} = 1$ SUSY getting enhanced to ${\cal N} = 3$ after an exactly marginal deformation we cannot have the sufficient condition satisfied. 
There also is a generalization of condition 4 that puts a restriction on ${\cal N} = 1$ index terms of the form $(-1)^{2j} t^{2+\tilde{R}} \chi_j(x)$ with $\frac{2j}{3} > \tilde{R} \geq \frac{2j-4}{3}$ for general $j$; it comes from the fact that only $\mathcal{S}$ multiplets in ${\cal N} = 1$ can contribute to these terms, and they pop up only in some special ${\cal N} = 3$ multiplets. However, this rule is very cumbersome to formulate, so we do not write it out explicitly. 
\paragraph*{Conditions with additional assumptions.}
Similarly to the previous subsection, if one makes the additional assumption (consistent with all known examples) that ${\cal N}=2$ theories do not have "spinning Coulomb branch operators" (namely $\mathcal{E}_{r,(0,j)}$ multiplets with $j>0$), then one can rule out existence of $\mathcal{B}_{[0; 0], R_3, (0, j)}$ multiplets for all $R_3$, $j$. Therefore, under this assumption for index to correspond to the pure ${\cal N} = 3$ theory, its corrected expansion should be 
\begin{equation}
	2t^{4/3} - 2t^{5/3}\left(x+\frac1x\right) + O(t^2)
\end{equation} 
and shall not contain any terms like
\begin{equation}
	t^{2+\tilde{R}} \chi_j(x); \qquad \frac{2j-2}{3} > \tilde{R} \geq \frac{2j-4}{3}.
\end{equation}

\subsection{SUSY enhancement and restrictions on ${\cal N}=1$ indices for enhancement to ${\cal N}=4$} \label{4req}
We list restrictions on the corrected index that can be useful for studying ${\cal N} = 1 \rightarrow {\cal N} = 4$ SUSY enhancement. Unlike for the ${\cal N} = 2, 3$ cases, the R-symmetry group does not have a $U(1)$, so the multiplet content of the theory is greatly restricted. For example, the only multiplet with bottom component scaling dimension less than 3 that can appear in an interacting ${\cal N} = 4$ theory is the stress-tensor multiplet. Using this knowledge, one can find the following restrictions:\\
{\bf Sufficient condition.} It is same as above.
\paragraph*{Necessary conditions.}
\begin{itemize}
		\item[1. ] The corrected ${\cal N} = 1$ index expansion for an interacting ${\cal N} = 4$ theory should look like 
		\begin{align}
			\mathcal{I}_4 &= 3n t^{2/3} -n t \left(x+\frac1x\right) + \left(6 -3n\right) t^{4/3} -3 t^{5/3}\left(x+1/x\right) \notag \\
			& + \left(10z+8y-7+2n\right) t^2  - \alpha t^{7/3} + \left(3-6z-9y\right) t^{7/3} \left(x+1/x\right) + O\left(t^{8/3}\right)
		\end{align}
		Here $n$ is the number of ${\cal N} = 4$ free field multiplets, $\alpha$ corresponds to the number of $\mathcal{C}^{1, 0}_{[0; 0; 0],(1/2, 0)}$ multiplets in the ${\cal N} = 4$ theory, $z$ to the number of $\mathcal{B}^{\frac12, \frac12}_{[0; 3; 0]}$ multiplets, and $y$ to the number of $\mathcal{B}^{\frac14, \frac14}_{[1; 1; 1]}$ multiplets\footnote{These multiplets are described in many details in \cite{Osborn}.}. This expression can be obtained by noting that the only ${\cal N} = 4$ multiplet that can contribute to $t^{7/3}$ is $\mathcal{C}^{1, 0}_{[0; 0; 0],(1/2, 0)}$, that $\mathcal{B}^{\frac12, \frac12}_{[0; 3; 0]}$ has in its ${\cal N} = 1$ decomposition 10 $\mathcal{S}_2$ and 6 $\mathcal{S}_{7/3, (0,\frac12)}$ multiplets (and the rest contribute as $O(t^{8/3})$), and that $\mathcal{B}^{\frac14, \frac14}_{[1; 1; 1]}$ has in its ${\cal N} = 1$ decomposition 8 $\mathcal{S}_2$ and 9 $\mathcal{S}_{7/3, (0,\frac12)}$ multiplets. We also used the fact that the ${\cal N} = 4$ stress-tensor multiplet decomposes into ${\cal N} = 1$ multiplets as 
		\begin{equation}
			\mathcal{B}^{\frac12, \frac12}_{[0; 2; 0]} \rightarrow 6 \mathcal{S}_{4/3} \oplus 8\hat{\mathcal{H}}_0 \oplus 3 \hat{\mathcal{H}}_{(0, \frac12)} \oplus 3 \hat{\mathcal{H}}_{(\frac12, 0)} \oplus 3 \mathcal{S}_{5/3, (0,\frac12)} \oplus \mathcal{S}_2 \oplus \hat{\mathcal{H}}_{(\frac12, \frac12)} 
		\end{equation}
		\item[2. ] Terms in the index must feature $t$ in powers of $\frac{n}{3}$ with $n$ integer. This is because the ${\cal N} = 4$ R-symmetry group does not have a $U(1)$ subgroup, so all operators in short representations have integer or half-integer scaling dimension.
		\item[3. ] If there is a term of the form $t^{2+\tilde{R}} \chi_j(x)$ with $\tilde{R} = \frac{2j-3}{3}$, then the theory can't have ${\cal N} = 4$ SUSY. This is because such a term can only lie in an ${\cal N} = 2$ $\mathcal{E}_{2\tilde{R}/3, (0, j)}$ multiplet, which cannot appear in a decomposition of an ${\cal N} = 4$ multiplet. Terms with $\tilde{R} = \frac{2j-4}{3}$ can appear in a free theory only.
\end{itemize}
One also can write an analogue of condition 3 for terms with $\frac{2j}{3} > \tilde{R} \geq \frac{2j-2}{3}$, but it turns out to be extremely cumbersome and not very useful. It is possible that there are more necessary conditions on ${\cal N} = 1$ index for enhancement to ${\cal N} = 4$, since the relation between ${\cal N} = 2$ and ${\cal N} = 1$ equivalence classes does not generalize easily for ${\cal N} = 4$; it would be interesting to work out the details of such a relation.

\section{AMS and SUSY enhancement} \label{ams}
In this section we will study theories that appear at the end of an RG flow, triggered by giving a nilpotent vev to a gauge singlet $M$ coupled to the moment map operator $\mu$ via $W = \mathrm{Tr} \mu M$; this idea, which we call in short ``AMS'' (for Agarwal, Maruyoshi, and Song, who wrote many papers on the subject) was introduced in \cite{GMTY} by Gadde, Maruyoshi, Tachikawa, and Yan and developed in \cite{Maruyoshi, Song, Agarwal, Sciarappa, Benvenuti, Giacomelli, Benvenuti2, Razamat}. A proper introduction into the subject can be found in \cite{Song}, and one particular example is studied in detail in \cite{Maruyoshi}, we will provide a short recap of ideas presented in \cite{Song} in the subsection below\footnote{A slightly different approach to the subject can be found in \cite{Benvenuti, Giacomelli}.}. In subsection \ref{checking} we will use the SUSY-enhancement checks we found above for several theories studied in \cite{Agarwal} and prove that they do not have ${\cal N} = 1 \rightarrow {\cal N} = 2$ SUSY enhancement.

\subsection{Description of the method} \label{method}
Let us assume we have a $d=4$ ${\cal N} = 2$ SCFT $\mathcal{T}$ with non-Abelian flavor symmetry $F'$, and R-symmetry $SU(2)_r \times U(1)_R$. Due to the presence of flavor symmetry the theory has a certain number of $\hat{\mathcal{B}}_1$ multiplets whose lowest component $\mu$ is a scalar that has charge 1 under the $SU(2)_r$ Cartan $I_3$ and is neutral under the $U(1)_R$ charge $R_2$. In agreement with the notation of \cite{Maruyoshi}, we introduce $J_+ = 2I_3$, $J_- = R_2$; $\mu$ has charges $(2, 0)$ under $(J_+, J_-)$, and $M$ has charges $(0, 2)$ (we will use this notation from now on). The whole procedure now goes on as follows: 

\begin{itemize}
	\item[1.] We deform $\mathcal{T}$ by adding an ${\cal N} = 1$ chiral multiplet $M$ transforming in the adjoint representation of $F$, and the superpotential coupling $W = \mathrm{Tr} \mu M$. This superpotential breaks the supersymmetry to ${\cal N} = 1$; the new R-symmetry is $R = \frac{2}{3}J_+ + \frac{1}{3}J_-$ and there also is residual symmetry $F_0 = \frac{1}{2}(J_+ - J_-)$ that becomes a global symmetry of the ${\cal N} = 1$ theory. This superpotential is by itself a marginally irrelevant term and in the IR theory simply decouples into the original one and the free chiral multiplets we added in the beginning.
	\item[2.] Now we give a nilpotent vev to $M$. According to the Jacobson-Morozov theorem (see e.g. chapter 3 in \cite{Collingwood} for an introduction into subject), any nilpotent element of a semi-simple Lie algebra $\mathfrak{f}$ is given via an embedding $\rho : \mathfrak{su}(2) \rightarrow \mathfrak{f}$ as $\rho(\sigma^+)$ (here and below $\sigma^+$, $\sigma_3$ correspond to the related Cartan subalgebra). Under such an embedding the adjoint representation of $\mathfrak{f}$ decomposes into $\mathrm{adj} \rightarrow \bigoplus\limits_j V_j \otimes R_j$, where $V_j$ is a spin-$j$ representation of $\mathfrak{su}(2)$ and $R_j$ is a representation of the residual flavor algebra $\mathfrak{h}$ under the embedding $\rho$. The vev breaks $J_-$, but preserves $J_- - 2\rho(\sigma^3)$. We denote the $SU(2)$ used in the embedding as $SU(2)_\rho$.

	For classical Lie algebras every embedding is specified by a certain partition of $N$ $[n_1, \ldots n_k]: n_i \geq n_{i+1}, \sum n_i = N$, where $N$ is defined by the flavor algebra $F'$. In a generic case not all partitions correspond to an embedding, we cover that in more details below; the systematic treatment of the problem can be found in \cite{Collingwood}. As a certain number $n_k$ can appear in the partition more than once, we introduce the shorthand notation for the partition $[n_1^{m_1}, \ldots n_k^{m_k}]$, where $m_i$ correspond to the number of times $n_i$ appears in the partition. Note that this is different from the notation in \cite{Agarwal}, our $n_k$ corresponds to their $k$ and our $m_k$ corresponds to their $n_k$. For example, one of the possible partitions of 13 is $[3,3,2,2,2,1]$ that can be rewritten in our notation as $[3^2, 2^3, 1]$. For exceptional Lie algebras there is a classification of nilpotent orbits in chapter 7 of \cite{Collingwood} and the adjoint decomposition for all $E_6$ and $E_7$ nilpotent orbits has been worked out in \cite{Chacaltana} and \cite{Chacaltana2}, respectively.
	\item[3.] After Higgsing, the superpotential becomes 
	\begin{equation}
		W = \mu_{1,-1,1} + \sum\limits_{j,j_3,f} M_{j,-j_3,f} \mu_{j,j_3,f}
	\end{equation}
	where $j$ is the spin from the adjoint decomposition, $j_3$ is the $\sigma_3$ eigenvalue and $f$ labels the representation under the new flavor symmetry group $F$. However, due to flavor current non-conservation the components of the superpotential with $j_3 \neq j$ combine with the current and become non-BPS; therefore the corresponding $M$ multiplets decouple. Thus we get that only $M$ multiplets of the $M_{j,-j,f}$ kind will stay, and they are coupled to $\mu_{j,j,f}$; the related superpotential is 
	\begin{equation}
		W = \sum\limits_{j,f} M_{j,-j,f} \mu_{j,j,f}
	\end{equation}
	\item[4.] If we start from a Lagrangian theory with matter fields in the fundamental rep of $F'$, the quark content of the theory can be read off from the partition. If the partition is $[n_1^{m_1}, \ldots n_k^{m_k}]$, then there are $m_1$ quarks that lie in the dimension-$n_1$ representation of $SU(2)_\rho$, \ldots, and $m_k$ quarks that lie in the dimension-$n_k$ representation of $SU(2)_\rho$. At the end of the RG flow they will have $(J_+, J_-)$ charges $(1, 1-n_1)$, $(1, 1-n_2)$, \ldots, $(1, 1-n_k)$.
	\item[5.] From the quark content one can simply calculate the meson content of the theory. Mesons lie in the adjoint of $F'$, and we can find the adjoint decomposition from the tensor product of 2 fundamental reps (or fundamental and antifundamental for $SU(N)$). For $SO$ ($Sp$) flavor group the tensor product will contain symmetric and antisymmetric part, and the adjoint will be antisymmetric (symmetric) respectively, so we will need to keep track of that.
	\item[6.] Breaking $J_-$ and adding the superpotential also affects the anomaly coefficients for $J_+$, $J_-$, so we need to recalculate them accordingly. Anomalies can be expressed in terms of meson operators according to \cite{Agarwal} and \cite{Maruyoshi} as
	\begin{align}
		\Tr J_+ & = \Tr J_+^3 = -N_M\\
		\Tr J_- &= 48(a_{UV}-c_{UV}) + \sum\limits_M \mathrm{d}(M) \\
		\Tr J_-^3 &= 48(a_{UV}-c_{UV}) - 6 I_y k_f + \sum\limits_M \mathrm{d}(M)^3\\
		\Tr J_+^2 J_- & = 8(2a_{UV}-c_{UV}) + \sum\limits_M \mathrm{d}(M)\\
		\Tr J_+ J_-^2 &= -  \sum\limits_M \mathrm{d}(M)^2
	\end{align}	
	Here $N_M$ is the total number of operators that got a vev and $\mathrm{d} (M_{-j, j}) = 2j+1$, $a_{UV}, c_{UV}$ correspond to the central charges in the UV, $k_f$ denotes the flavor central charge, sums are taken over $M_{j,-j}$, and $I_y$ is the embedding index.
	\item[7.] The R-symmetry in the IR will be given by some mixture of $J_+, J_-$: $R_{IR} (\epsilon) = \frac{1+\epsilon}{2} J_+ + \frac{1-\epsilon}{2} J_-$ (we assume there is no global accidental symmetry in the IR). The exact value of epsilon can be determined by $a$-maximization \cite{Intriligator}: we need to maximize 
	\begin{equation}
		a(\epsilon) = \frac{9}{32} \Tr R_{IR}(\epsilon)^3 - \frac{3}{32} \Tr R_{IR}(\epsilon)
	\end{equation}
	\item[8.] After doing a-maximization we need to check whether all scalar chiral operators satisfy unitarity bounds (whether they have $R_{IR} \geq \frac{2}{3}$). If this is not the case for some of the operators, this means that they become free and decouple. Therefore, we need to subtract them according to \cite{Kutasov} (note that in \cite{Kutasov} $\tilde{a}$ is defined as $\frac{3}{32}a$) and maximize the new $a$:
 	\begin{equation}
 		a_n(\epsilon) = a_{n-1} (\epsilon) + \frac{1}{96}\sum\limits_M \#(M) \left[(2-3R(M))^2 (5-3R(M))-2\right], \label{arecorrected}
 	\end{equation}
 	$a_{n-1}$ is the current trial $a$, $M$ runs over operators with different R-charges such that $R(M)\leq\frac{2}{3}$, $\#(M)$ is the number of operators with given R-charge and $R(M)$ is the R-charge under the current trial R-symmetry. One should notice that our expression disagrees with \cite{Kutasov} by a factor of $\frac{1}{48}$ that corresponds to subtraction of an ${\cal N} = 1$ multiplet (that is coupled to $M$) from the central charge. After doing the subtraction we should recalculate $\epsilon$ and check whether any new operators hit the unitarity bound; this should be repeated until all operators in the theory have $R(M) \geq \frac{2}{3}$. 

 	Apart from the usual suspects of $M_{j,f}$ we introduced in 3., operators of the kind $\Tr \phi^k$ (where $\phi$ is a chiral superfield in the adjoint of the gauge group), that are related to the Coulomb branch can also decouple (see \cite{Maruyoshi}). It is also possible that the quark operators $q^n$, $q^n \phi^k$ will decouple; however, gauge-invariant quark operators should be subtracted only if they can't be expressed as a derivative of the superpotential. In fact, in all cases we considered there was no need to subtract quark operators; we are not sure whether this is always the case. 
	\item[9.] $c$ can be computed in a similar fashion; $c$ is given by
	\begin{equation}
		c(\epsilon) = \frac{9}{32} \Tr R_{IR}(\epsilon)^3 - \frac{5}{32} \Tr R_{IR}(\epsilon),
	\end{equation}
	and subtracting decoupled operators goes on as 
	\begin{equation}
		c_n(\epsilon) = c_{n-1} (\epsilon) + \sum\limits_M \#(M) [\frac{1}{96}(2-3R(M))(8-21R(M)+9R(M)^2) - \frac{1}{24}].
	\end{equation}	
	After all these manipulations we obtain the correct $a$, $c$ and R-symmetry. 
	\item[10.] We can now compute the ${\cal N} = 1$ index of the resulting theory. For example, the index of a theory that lies at the end of the RG flow of SQCD with $SU(N)$ gauge and $SU(2N)$ flavor group for a general partition and no quark operator decoupling is given by 
	\begin{align}
		\mathcal{I}\left(t,x,\xi\right) = &\frac{\kappa^{N-1}}{N!} \frac{\prod\limits_M \Gamma\left(t^{\left(1+j\right)\left(1-\epsilon\right)} \xi^{-\left(1+j\right)}\right)}{\prod\limits_{\phi^k} \Gamma\left(t^{2k\left(1-\epsilon\right)} \xi^{-2k}\right)} \Gamma\left(\frac{t^{1-\epsilon}}{\xi}\right)^{N-1} \oint \prod\limits_{i=1}^{N-1} \frac{d z_i}{2\pi i z_i} \label{gammaindex} \notag \\  & \prod\limits_{\alpha \in \Delta} \frac{\Gamma\left(\mathbf{z}^\alpha \frac{t^{1-\epsilon}}{\xi}\right)}{\Gamma\left(\mathbf{z}^\alpha\right)}  \prod\limits_{w \in R} \Gamma\left(\mathbf{z}^{\pm w} t^{J^q_+\left(1+\epsilon\right)/2+J^q_-\left(1-\epsilon\right)/2}\xi^{\left(J^q_+ - J^q_-\right)/2}\right) 
	\end{align}

	Here $t$ and $x$ are the usual ${\cal N} = 1$ index fugacities, $\xi$ is a fugacity for the residual ${\cal N} = 1$ flavor symmetry, $$\kappa = (tx; tx) (t/x; t/x) = \prod\limits_{n>0} (1-(tx)^n)(1-(t/x)^n);$$ $\frac{\kappa^{N-1}}{N!}$ comes from the $SU(N)$ gauge group. $\Gamma(q)$ is a shorthand for the elliptic gamma function $$\Gamma(q) \equiv \Gamma(q, tx, t/x) = \prod\limits_{m,n\geq 0} \frac{1-q^{-1}t^{m+n+2}x^{m-n}}{1-qt^{m+n}x^{m-n}}.$$ The product
	$$\prod\limits_M \Gamma(t^{(2+2j)(1-\epsilon)} \xi^{-(2+2j)})$$ is taken over all $M$ that were added in the superpotential and did not decouple. $$\prod\limits_{\phi^k} \Gamma(t^{2k(1-\epsilon)} \xi^{-2k})$$ corresponds to the $\Tr \phi^k$ operators that decoupled. $\Gamma(\frac{t^{1-\epsilon}}{\xi})^{N-1}$ corresponds to the zero roots of $SU(N)$, the product $\prod\limits_{\alpha \in \Delta}$ is taken over non-zero roots of $SU(N)$. $$\prod\limits_{\alpha \in \Delta} \frac{\Gamma(\mathbf{z}^\alpha \frac{t^{1-\epsilon}}{\xi})}{\Gamma(\mathbf{z}^\alpha)}$$ corresponds to the vector multiplets contribution; $\mathbf{z}^\alpha$ should be read as $\prod z_i^{\alpha_i}$. The notation $\Gamma(z^{\pm a} x)$ is a shorthand for $\Gamma(z^{a} x) \Gamma(z^{-a} x)$, $\prod\limits_{w \in R}$ is taken over the weights of the fundamental representation of $SU(N)$ and $J^q_+, J^q_-$ are quark charges under $J_+, J_-$. The expression is altered slightly for different gauge groups. 

	\item[11.] Unfortunately, straightforward index calculation is very computationally involved with complexity growing for the larger groups (so Mathematica is unable to compute relevant index terms for $SO(20)$ and larger groups). An alternative is to recall the definition of the index through the plethystic exponential. Using this definition, we can rewrite the index as 
	\begin{align}
		\mathcal{I} &= \frac{\prod\limits_M \Gamma(t^{(1+j)(1-\epsilon)} \xi^{-(1+j)})}{\prod\limits_{\phi^k} \Gamma(t^{2k(1-\epsilon)} \xi^{-2k})} \int d \mu (g)\exp \left(\frac1n \sum\limits_{n = 1}^\infty \left[\sum\limits_q I_q(t^{n}, x^n, g^n, f^n) + \notag \right.  \right. \\ 
		& + I_\phi(t^{n}, x^n, g^n) + I_g(t^{n}, x^n, g^n)\bigg]\bigg)
	\end{align}
	where the prefactors are as in \eqref{gammaindex}, and the integral is taken over the gauge group. In the exponent $I_q$ corresponds to the quark contribution, and $I_\phi$, $I_g$ are related to the gluon contribution; they are given by
	\begin{align}
		I_q & = \frac{t^{1-(1-\epsilon)r/2} - t^{1+(1-\epsilon)r/2}}{(1-t/x)(1-tx)} \chi_f (g) \chi_f (f)\\
		I_\phi & = \frac{t^{1-\epsilon} - t^{1+\epsilon}}{(1-t/x)(1-tx)} \chi_a (g) \\
		I_g & = \frac{2t^{2} - t\left(x+\frac1x\right)}{(1-t/x)(1-tx)} \chi_a (g)
	\end{align}
	where $\chi_f (g)$ is the character for the fundamental representation of the gauge group, $\chi_f (f)$ is the character for the fundamental representation of the quark flavor group (and $f$ as an overall prefactor corresponds to the fundamental rep of some group), $\chi_a (g)$ is the character for the adjoint representation of the gauge group (and $a$ as an overall prefactor corresponds to the adjoint rep), and $r$ corresponds to the quark representation under $SU(2)_\rho$. One can use the formulas from \cite{Dolan} to see that the integration gives same result as \eqref{gammaindex}.

	To proceed with the computation, we can use:
	\begin{equation}
		\sum_{n= 0}^{\infty}t^n\chi_{S^n_\alpha}(g)=\exp\left\{\sum_{p=1}^{\infty}t^p\frac{\chi_\alpha(g^p)}{p}\right\},
	\end{equation}
	where $S^n_\alpha(g)$ is n-th symmetric power of the representation $\alpha$ in the group $g$. This strongly reminds us of the expressions for the index and allows us to express the contributions of $I_\phi$, $I_g$ in a simpler form. The quark contributions are, however, harder to account for; e.g. the contribution at order $f^2, g^2$ looks like $\frac{1}{2}(\chi_f(f^2) \chi_f(g^2) + \chi_f(f)^2 \chi_f(g)^2)$ which does not factorize into $S^2_f(f) S^2_f(g)$. Such terms can be expressed as a sum of plethysms, but in cases we encountered it was often easier to handle these cases manually. For brevity we will denote them as $S^n_f(f) \otimes S^n_f(g)$.
\end{itemize}

As an illustration of the method we described above, let us analyze a theory that was studied in detail in \cite{Maruyoshi} and discuss it thoroughly.
Let us start with an ${\cal N} = 2$ supersymmetric $SU(2)$ gauge theory with $N_f = 4$ fundamental hypermultiplets and $SO(8)$ global symmetry, and add to it a superpotential $W= \Tr M\mu$. Next we give a nilpotent vev to $M$ by $\rho(\sigma^+)$; $\rho: \mathfrak{su}(2) \rightarrow \mathfrak{so}(8)$ that is described by a partition $[7; 1]$. This means that we fully break the flavor symmetry and the adjoint of $SO(8)$ decomposes as $\mathbf{28} \rightarrow V_1 \oplus V_3 \oplus V_3 \oplus V_5$ (here $V_j$ correspond to the spin-$j$ representation of $SU(2)$) and we get operators $M_{j,-j}$ with $j=1,3,3,5$ and $(J_+, J_-)$ charges $(0,4)$, $(0,8)$, $(0,8)$, $(0,12)$ respectively. The anomaly coefficients after the deformation are given by 
\begin{align}
	&\Tr J_+ = \Tr J_+^3 = -4, \; \Tr J_- = 18, \; \Tr J_- = 18, \notag \\
	&\Tr J_-^3 = 1362, \; \Tr J_+^2 J_- = 34,\; \Tr J_+ J_-^2 = -228. \notag
\end{align}
This gives us a trial $a$-charge $$a(\epsilon) = -\frac{3}{32} (807 \epsilon^3 - 1746 \epsilon^2 + 1231\epsilon -284),$$ that after $a$-maximization gives us $\epsilon = \frac{582+\sqrt{7585}}{807} \approx 0.83$. This means that $\Tr \phi^2$ and $M_{1,-1}$ decouple and according to \eqref{arecorrected} we get $$a_1 (\epsilon) = -\frac{1}{96}(2520 - 10791 \epsilon + 15066 \epsilon^2 - 6831 \epsilon^3).$$ Re-maximizing new $a$ yields a new $\epsilon = \frac{558+\sqrt{8017}}{759}$, that signifies decoupling of two $M_{-3,3}$ operators, and finally a second re-maximization yields $\epsilon = \frac{13}{15}$, $a=\frac{43}{120}$, and $c=\frac{11}{30}$. The value of $2a-c$ from the central charges and from the Coulomb branch operators according to the Tachikawa-Shapere formula \cite{Shapere} match, consistent with a possible enhancement of SUSY to ${\cal N} = 2$.

\subsection{SUSY enhancement checking} \label{checking}
Using the index computation procedure we have derived in subsection \ref{method}, and the SUSY enhancement conditions from section \ref{req}, we were able to compute the following index series expansions for different theories studied by \cite{Agarwal}. The theories in question lie at the end of an RG flow from ${\cal N}=2$ SQCDs with a flavor group $F$:
\begin{itemize}
	\item[1.] $F=Sp(2)$ with $[2, 1^2]$ partition. The theory has $SU(2) \times U(1)$ symmetry in the IR, and the corrected index expansion is given by
	\begin{align}
		\mathcal{I} = &2 t^{16/21}+3 t^{32/21}+3 t^{64/63} +3 t^{94/63} -2 t^{95/63}\left(x+\frac1x\right)+2 t^{110/63} \notag - \\ 
		& -2 t^2 + 6 t^{16/9} +6 t^{128/63} - 2 t^{47/21}\left(x+\frac1x\right)+4 t^{16/7}+2 t^{142/63} \notag - \\
		& -4 t^{143/63}\left(x+\frac1x\right) -6 t^{52/21}+ O(t^{\frac73})
	\end{align}
	One can notice the absence of a $t^\frac{4}{3}$ term, even though the corresponding coefficient according to section \ref{req} should be at least $(3+1)-3=1$ for the SUSY enhancement to happen. Therefore, there is no SUSY enhancement in this case.	
	\item[2.] $F=SO(8)$ with $[5, 3]$ partition. The theory has $U(1)$ symmetry in the IR, and the corrected index expansion is given by
	\begin{align}
		\mathcal{I} = &2 t^{40/51}-t^{61/51}\left(x+\frac1x\right)+t^{72/51}+3 t^{80/51}+2 t^{82/51}-t^{92/51} -\notag \\
		&-2 t^{101/51} \left(x+\frac1x\right)-3 t^2+ O(t^{\frac73})
	\end{align}
	Let us notice the term $$-t^{61/51}\left(x+\frac1x\right).$$ If we have ${\cal N} = 1 \rightarrow {\cal N} = 2$ SUSY enhancement, then it should lie in the ${\cal N} = 2$ $\mathcal{E}$ multiplet. It cannot lie in an $\mathcal{E}$ multiplet with scalar bottom component, as there is no $t^{\frac{44}{51}}$ term in the index. Thus it should be in the bottom component of $\mathcal{E}$. As $\mathcal{E}$ obey the decomposition rule \eqref{emult}, an ${\cal N} = 1$ multiplet $\mathcal{S}_{\frac{78}{51},(0, 1)}$ should be also present in the theory and contribute to the index as $$t^{78/51}(x^2+1+\frac{1}{x^2}).$$ This contribution cannot be canceled by contributions of another multiplets and is not present. Therefore, this theory does not exhibit ${\cal N} = 1 \rightarrow {\cal N} = 2$ SUSY enhancement.
	\item[3.] $F=SO(8)$ with $[7, 1]$ partition (the case discussed in section \ref{method}). The theory has $U(1)$ symmetry in the IR, and the corrected index expansion is given by
	\begin{align}
		\mathcal{I} & = t^{4/5} - t^{17/15}\left(x+\frac1x\right) + t^{22/15} + t^{8/5} - t^{29/15}\left(x+\frac1x\right) - t^2 \notag + \\ 
		& + t^{34/15} + t^{7/3}\left(x+\frac1x\right) + O(t^{12/5})
	\end{align}
	One can notice that unlike in the previous case, the low-$t$ terms combine into the contribution of ${\cal N} = 2$ $\mathcal{E}$ multiplet 
	\begin{equation}
		t^{4/5} - t^{17/15}\left(x+\frac1x\right) + t^{22/15} \leftrightarrow \mathcal{E}_{6/5, (0,0)},
	\end{equation}
	so the necessary conditions are satisfied, and we got the Coulomb branch operator $\mathcal{E}_{6/5, (0,0)}$. Moreover, there is a term $$t^{7/3}\left(x+\frac1x\right)$$ that signals the presence of a conserved supercurrent and SUSY is enhanced to ${\cal N} = 2$. Indeed, the obtained index correspond to the $(A_1, A_2)$ Argyres-Douglas theory, which was first shown by Maruyoshi and Song in \cite{Maruyoshi}. The $U(1)$ conserved current becomes part of the R-symmetry, so there are no conserved current multiplets in the theory.
	\item[4.] $F=SO(20)$ with $[3^4, 2^4]$ partition. Here Mathematica fails to yield the answer, so one has to resort to computations through plethystic exponentials. The theory has $SO(4) \otimes Sp(2) \otimes U(1)$ symmetry in the IR and the gauge group is $Sp(4)$ (in our notation $Sp(2)$ has dimension 10), and from a-maximization we find $\epsilon = \frac{2}{3}$. The quark contributions to the index are given by
	\begin{align}
		I_3 &= \frac{t^{1/2} - t^{3/2}}{(1-t/x)(1-tx)} \chi_{11}(z_1, z_2) \xi_{1000} (g_1, g_2, g_3, g_4) \\
		I_2 &= \frac{t^{2/3} - t^{4/3}}{(1-t/x)(1-tx)} y_{10}(y_1, y_2) \xi_{1000} (g_1, g_2, g_3, g_4)
	\end{align}
	Here we denote $Sp(2)$ characters with $y_{rep}$, $SO(4)$ characters with $\chi_{rep}$ (so the fundamental has character $\chi_{11} (z_1, z_2) = (z_1 + \frac{1}{z_1})(z_2 + \frac{1}{z_2})$) and $Sp(4)$ characters with $\xi_{rep}$. Using this knowledge, we can write how $I_3$, $I_2$, $I_\phi$, $I_g$ contribute to the index up to order $t^{2}$; we will drop indices for the fundamental reps and denote adjoint of the $Sp(4)$ with $\xi_a$ for brevity:
	\begin{align}
		I_3 & = 1 + t^{1/2} \chi \xi + t S^2\left(\chi\right) \otimes S^2\left(\xi\right) + t^{3/2} \left(S^3\left(\chi\right) \otimes S^3\left(\xi\right) +\left(x+\frac1x - 1\right)\chi \xi\right) + \notag \\ 
		&+ t^2 \left(S^4\left(\chi\right) \otimes S^4\left(\xi\right) +\left(x+\frac1x\right)\chi \xi - \chi^2 \xi^2\right) + O\left(t^{5/2}\right) \\
		I_2 & =1 + t^{2/3} \xi y + t^{4/3}\left(S^2\left(y\right) \otimes S^2\left(\xi\right) - \xi y\right) + t^{5/3} \xi y\left(x+\frac1x\right) + \notag\\ 
		&+ t^{2}\left(S^3\left(y\right) \otimes S^3\left(\xi\right)- \chi^2 \xi^2\right)  + O\left(t^{8/3}\right) \\
		I_g &= 1 - t\left(x+\frac1x\right) \xi_a + t^2 \{S^2\left(\xi_a\right)+\xi_a+\notag \\ 
		&+\left(x^2+1+\frac{1}{x^2}\right)\left[\xi_a^2 - \xi_a - S^2\left(\xi_a\right)\right]\} + O\left(t^{3}\right) \\ 
		I_\phi &= 1 + t^{1/3} \xi_a + t^{2/3} S^2\left(\xi_a\right) + t S^3\left(\xi_a\right) + t^{4/3}\left(S^4\left(\xi_a\right) +\left(x+\frac1x\right)\xi_a\right) + \notag \\
		&+ t^{5/3} \left[S^5\left(\xi_a\right) + 2\left(x+\frac1x\right)S^2\left(\xi_a\right) - \xi_a\right] +t^2 \left[S^6\left(\xi_a\right)-\notag \right. \\ 
		&- \left. \xi_a^2+\left(x+\frac1x\right)\{\xi_a S^2\left(\xi_a\right)\}\right] 
		+ O\left(t^{7/3}\right)
	\end{align}

	We can also write series expansion for the prefactor contribution: 
	\begin{align}
		\frac{\Gamma(t^{5/6})^{16} \Gamma(t)^{6}}{\Gamma(t^{2/3})} & = 1 - t^{2/3} + 16 t^{5/6} - 16 t^{7/6} + t^{4/3} - 16 t^{3/2} + \notag \\ &+ t^{5/3} \left(-x - \frac1x + 136\right)+ 16t^{11/6}\left(x+\frac1x\right) - 257t^2+ O(t^{13/6})
	\end{align}
	The next thing one should do is to take the product of all these terms and leave only singlets of $Sp(4)$; this is a very tedious computation that can be partially simplified by using LiE \cite{Lie}; after a lengthy computation one gets the corrected index
	\begin{align}
		\mathcal{I} &= 16 t^{5/6} + 6 t +t^{4/3}\left(-x -\frac1x + 18\right) + 16 t^{3/2} + 152 t^{5/3} +96 t^{11/6} + \notag\\ 
		&+ t^2 \left(6-5\left(x+\frac1x\right)\right) + O(t^{13/6})
	\end{align}
	The terms up to $t^{11/6}$ have been computed independently through a gamma-function expansion; the results agree. One can see from the expression that ${\cal N} = 1 \rightarrow {\cal N} = 2$ enhancement cannot happen: we have a term $16 t^{5/6}$ that can lie only in $\mathcal{E}_{5/4, (0,0)}$. However, the coefficient in front of $$t^{7/6}\left(x+\frac1x\right)$$ is zero, while if we would have an ${\cal N} = 2$ theory with 16 $\mathcal{E}_{5/4, (0,0)}$ multiplets, it would have been $-16$ or less. Therefore, no enhancement can happen.
	\item[5. ] $F=SO(20)$ with $[9, 5, 3, 1^3]$ partition. Once again, we need to do the computations using the plethystic exponential. $\epsilon = \frac{5}{6}$, mesonic prefactor is 
	\begin{equation}
		\frac{\Gamma(t^{4/3})\Gamma(t^{7/6})^8\Gamma(t)^3\Gamma(t^{5/6})^8}{\Gamma(t^{1/3})},
	\end{equation}
	and index expansion in $t$ yields
	\begin{equation}
		\mathcal{I} = 5t^{5/6}+4t-t^{7/6}\left(x+\frac1x\right)+t^{7/6}+8t^{4/3}+O(t^{3/2}).
	\end{equation}
	One can see that once again there is no SUSY enhancement, as the term $5 t^{5/6}$ can lie only in $\mathcal{E}_{5/4, (0,0)}$ multiplet, and we should also have a term $\alpha t^{7/6}\left(x+\frac1x\right)$ with $\alpha<-5$, which is absent.
	\item[6. ] $F=SO(24)$ with $[5, 2^8, 1^3]$ partition. Once again, we need to do the computations using the plethystic exponential. $\epsilon = \frac{2}{3}$, mesonic prefactor is 
	\begin{equation}
		\frac{\Gamma(t^{4/3})\Gamma(t^{7/6})\Gamma(t)^3\Gamma(t^{5/6})^5}{\Gamma(t^{1/3})\Gamma(t^{2/3})},
	\end{equation}
	and index expansion in $t$ yields
	\begin{equation}
		\mathcal{I} = 8t^{5/6}+3t+8t^{7/6}+t^{4/3}\left(32-x-\frac1x\right)+O(t^{3/2}).
	\end{equation}
	One can see that once again there is no SUSY enhancement, as term $8 t^{5/6}$ can lie only in $\mathcal{E}_{5/4, (0,0)}$ and we should also have a term $\alpha t^{7/6}\left(x+\frac1x\right)$ with $\alpha<-8$, which is absent.
\end{itemize}

\section{Index and SUSY enhancement conditions in 3d} \label{d3}
Similarly to the $d=4$ case, all conditions can be divided into necessary, sufficient and sanity checks. A protected supercurrent multiplet (and, therefore, a necessary and sufficient condition) appears only in ${\cal N}>3$ multiplets and in these cases it is very easy to check whether there is SUSY enhancement. For example, let us define ${\cal N}=4$ index as
\begin{equation}
	\tilde{\mathcal{I}}_{4} = \Tr{(-1)^F x^{\Delta+j} y^{R_1-R_2}},
\end{equation}
where the trace is taken over all Verma module states (but only those that lie in the kernel of $\delta$ contribute); $\delta = \frac12 \{Q^\dagger, Q\}$, $Q$ is a specific supercharge, that has $SO(4)_R$ charge $(1; 1)$ and spin $-\frac12$, and $\Delta$ is the scaling dimension. Under such a definition for SUSY to be enhanced from ${\cal N}=4$ to ${\cal N}=4+\alpha$ the coefficient in front of $x y^0$ term (that is related to supercurrent multiplet $B^{(1; 1)}$) in the index series expansion should be $\alpha$ and vice versa.

\subsection{${\cal N}=1$ index}\label{3dn1}
$3d$ ${\cal N}=1$ superconformal theories are very special, because ${\cal N}=1$ superconformal algebra does not contain any R-symmetry. The only short multiplets are the free field multiplets and conserved current multiplets; superconformal index looks more like a Witten index and is a number $I$ that, after taking free fields into account, tracks the difference between the number of conserved fermionic and conserved bosonic currents. If one assumes there is no free sector and only one stress-tensor in the theory, then using the fact that the only conserved fermionic current allowed in non-free theory is the supercurrent, one can derive necessary condition on $I$ to possibly correspond to ${\cal N}>1$ SUSY; this condition is very weak and is of very doubtful usefulness, but we list it for the sake of completeness. It turns out that $I\geq1$ for theory to be able to have ${\cal N}=2$ (with its value depending on the dimension of the flavor symmetry group), $I\geq 2$ for possibility of enhancement to ${\cal N}=3$, $I\geq4$ for possible enhancement to ${\cal N}=4$. For ${\cal N}>4$ we can use the fact, that ${\cal N}>4$ algebras cannot have conserved flavor currents and fix $I$ precisely: for possible enhancement to ${\cal N}=5$ $I$ should be equal to 6, for ${\cal N}=6$ $I=10$ and for ${\cal N}=8$ $I=21$. 

\subsection{${\cal N}=2$}
Proper discussion and derivation can be found in the appendix \ref{n2cond}, here we will simply define basic concepts and list the results. For ${\cal N}=2$ theories one can define index as $$\tilde{I}(x) = \Tr{(-1)^F x^{\Delta+j}},$$ where $j$ is the $SO(3)$ spin; all ${\cal N}=2$ multiplets contribute to the index as $$\pm \frac{x^\alpha}{(1-x^2)}.$$ Therefore, one can consider corrected index 
\begin{equation} 
	I = (1-x^2)(\tilde{I} -1)= \sum_k a_k x^k
\end{equation} 
($a$'s are the coefficients). The corrected index will be, loosely speaking, a power series in the scaling dimensions of superconformal multiplets bottom components. If we disregard free fields contribution, lowest in $x$ term should have $x$ in power greater than $\frac12$; for any SUSY enhancement to happen we need index contain only terms with $k$ integer or half-integer. There are no sufficient conditions on SUSY enhancement, as the supercurrent multiplet contribution to the index is same to the contribution of another multiplet. The necessary conditions on the coefficients $a_k$ of the corrected index for different cases are as follows:
\begin{itemize}
	\item[${\cal N}=3$.] In this case $a_1 +a_2 + 2 \geq 0$ and $a_1$ should be equal to the dimension of the flavor group in the enhanced-SUSY theory.
	\item[${\cal N}=4$.] In this case $a_1 +a_2 + 5 \geq 0$ and $a_1$ should be equal to the dimension of the flavor group in the enhanced-SUSY theory.
	\item[${\cal N}=5$.] In this case $a_1 = 1$, $a_2 \geq -3$ and $a_k$ should be even for non-integer $k$.
	\item[${\cal N}=6$.] In this case $a_1 = 4$, $a_2 \geq -9$ and $a_k$ should be even for non-integer $k$.
	\item[${\cal N}=8$.]\footnote{Any ${\cal N}=7$ SCFT with stress-tensor must have ${\cal N}=8$ SUSY \cite{Bashkirov}.}In this case $a_1 = 10$, $a_2 \geq -15$ and $a_k$ should be divisible by 4 for non-integer $k$.
\end{itemize}

\subsection{${\cal N}=3$}
${\cal N}=3$ index uses the same fugacities as ${\cal N}=2$ does, so one can simply consider contributions of ${\cal N}=3$ multiplets to the ${\cal N}=2$ index and derive from them the conditions on SUSY enhancement \ref{n3cond}. Most of the conditions are similar to ${\cal N}=2$ case, but there is a sufficient enhancement condition: 

If in the corrected index $I = \sum_k a_k x^k$ the coefficient $-a_2$ is greater than $a_1$, then there is ${\cal N}=3\rightarrow {\cal N}=3-a_2-a_1$ SUSY enhancement.

The necessary conditions on the coefficients $a_k$ of the corrected index for different cases are as follows (derivation can be found in the appendix \ref{n3cond}):
\begin{itemize}
	\item[${\cal N}=4$.] In this case the dimension of the flavor group in the enhanced-SUSY theory should be either $a_1-1$ or $a_1$.
	\item[${\cal N}=5$.] In this case $a_1 = 1$, $a_2 \geq -3$ and $a_k$ should be even for non-integer $k$.
	\item[${\cal N}=6$.] In this case $a_1 = 4$, $a_2 \geq -9$ and $a_k$ should be even for non-integer $k$.
	\item[${\cal N}=8$.] In this case $a_1 = 10$, $a_2 \geq -15$ and $a_k$ should be divisible by 4 for non-integer $k$.
\end{itemize}

\section{Index and SUSY enhancement conditions in 6d} \label{d6}
In 6 dimensions there are two superconformal algebras that admit SCFT: ${\cal N}=(1,0)$ and ${\cal N}=(2,0)$, the rotation group is $Spin(6)$ and R-symmetry is $Sp(1)$, $Sp(2)$ respectively. As ${\cal N}=(1,0)$ supercurrent multiplet is protected, its contribution to the index cannot be canceled, so there is a necessary and sufficient condition on SUSY enhancement from ${\cal N}=(1,0)$ to ${\cal N}=(2,0)$. Let us define ${\cal N}=(1,0)$ index (first suggested in \cite{Bhattacharya}, our notation is in agreement with \cite{Kim}) as 
\begin{equation}
	\tilde{\mathcal{I}}_6 = \Tr{(-1)^F q^{3R+j_1+(j_2+3j_3)/2}s^{j_2}}, 
\end{equation}
where $R$ is $Sp(1)_R = SU(2)_R$ Cartan, $j_1, j_2,j_3$ define $[j_1, j_2, j_3]$ representation of $SO(6)$ Lorentz group (so the supercharges $Q$ lie in $[0,1,0]$ representation and have R-charge $\pm\frac{1}{2}$), trace is taken over all operators, but only those lying in the kernel of $\mathcal{H}$ contribute; $\mathcal{H} = \frac12 \{Q^\dagger, Q\}$, where $Q$ is a specific supercharge with $SO(6)$ charge $[0,0,-1]$ and $SU(2)_R$ charge $\frac12$. Then the supercurrent multiplet contributes to the index as 
\begin{equation}
	\frac{-q^{7/2}(s+\frac1s+1)}{(1-q)(1-qs)(1-q/s)},
\end{equation}
and one can consider corrected index 
\begin{equation}
\mathcal{I} = (1-q)(1-qs)(1-q/s)\tilde{\mathcal{I}}_6
\end{equation}
and look for a $-q^{7/2}$ term (or $-q^{7/2} s$ or $-q^{7/2}/s$, as all three monomials are on the equal footing in the index expansion). If it is present, there is ${\cal N}=(1,0)$ to ${\cal N}=(2,0)$ SUSY enhancement, and if it's absent, there is no SUSY enhancement.

\bigskip
\noindent{\bf Acknowledgments}

I would like to thank Shlomo Razamat, Gabi Zafrir, Petr Kravchuk, Mikhail Isachenkov, Lorenzo Di Pietro, Ran Yacoby, and Alexander Tumanov for useful discussions, Zohar Komargodski for useful discussions and suggesting the idea of the project, and especially Ofer Aharony for useful discussions, suggesting the idea of the project, general guidance and comments on a draft of this manuscript.

\appendix

\section{Index and SUSY enhancement conditions derivation in 4 dimensions} \label{d4comp}
In this appendix we present various calculations and definitions, that are used in section \ref{enh}. In subsection \ref{multred} we describe, how various ${\cal N} = 2$ superconformal multiplets decompose into ${\cal N} = 1$ ones, and how ${\cal N} = 1$ supermultiplets contribute to the index. In subsection \ref{alg} we explain, how one can build relation between $d=4$ ${\cal N} > 1$ and ${\cal N} = 1$ supermultiplets, in the \ref{algN2} subsection we build it explicitly for ${\cal N} =2 \leftrightarrow {\cal N} = 1$ multiplets, and in the \ref{flavN2} we build it explicitly for ${\cal N} =2 \leftrightarrow {\cal N} = 1$ multiplets, when flavor symmetry is lifted to the R-symmetry.
\subsection{${\cal N} = 2$ to ${\cal N} = 1$ multiplet decomposition and index} \label{multred}
In this section we will list how all ${\cal N} = 2$ superconformal multiplets decompose into ${\cal N} = 1$ multiplets. Our notation for ${\cal N} = 2$ multiplets is as in \cite{Osborn}. For ${\cal N} = 1$ multiplets we have a notation of our own, that is somewhat similar to the one used in \cite{Cordova} by Cordova et al., and the relations are written out below. We denote ${\cal N} = 1$ long multiplet as $\mathcal{L} \leftrightarrow L\bar{L}$ (on the left side we use our notation and on the right --- the one used in \cite{Cordova}; please note that our definition of $U(1)_R$ charge has different sign cf. the one in \cite{Cordova}). The multiplets with semi-BPS shortening under $Q$ ($\bar{Q}$) are denoted as $\mathcal{H}_{R_1, (j, \bar{j})} \leftrightarrow A\bar{L}$\footnote{In section 4.5 of \cite{Cordova} one can find tables of all ${\cal N} = 1$ multiplets.} ($\bar{\mathcal{H}} \leftrightarrow L\bar{A}$). The multiplets with symmetric semi-BPS shortening are denoted as $\hat{\mathcal{H}}_{(j,\bar{j})} \leftrightarrow A\bar{A}$. Multiplets with BPS shortening are denoted as $\mathcal{S}_{R_1, (0, \bar{j})} \leftrightarrow B\bar{L}$. Finally, free field multiplets are denoted as $\hat{\mathcal{S}} \leftrightarrow B\bar{A}$, $\bar{\hat{\mathcal{S}}} \leftrightarrow A\bar{B}$. We drop spin charges for scalars for brevity. As in the general case there will be many long multiplets in the decomposition, we will not bother to write out the $R$-charge and spins for them precisely. We start with the general case and then discuss special cases where the general formulas can't be formally applied. In the last subsection \ref{n1index} we list the ${\cal N} = 1$ multiplets contributions to the index for reference.

\subsubsection{General case}
Moving from smaller to bigger, the ${\cal N} = 2$  multiplets decompose into ${\cal N} = 1$ multiplets as follows:
\begin{align}
	&\hat{\mathcal{B}}_r  \rightarrow \mathcal{S}_{\frac{4r}{3}} \oplus \bar{\mathcal{S}}_{\frac{-4r}{3}} \oplus \bar{\mathcal{H}}_{\frac{-4(r-1)}{3}} \oplus \mathcal{H}_{\frac{4(r-1)}{3}} \oplus (2r-3) \mathcal{L} \label{bhatmult}\\
	&\mathcal{E}_{\frac{R_2}{2},(0,\bar{j})}  \rightarrow \mathcal{S}_{\frac{2R_2}{3},(0,\bar{j})}\oplus\mathcal{S}_{\frac{2R_2+1}{3},(0,\bar{j}\pm\frac{1}{2})}\oplus \mathcal{S}_{\frac{2R_2+2}{3},(0,\bar{j})}\label{emult}
\end{align}
\begin{align}
	\mathcal{D}_{r,(0,\bar{j})} & \rightarrow \mathcal{S}_{\frac{4r+2+2\bar{j}}{3},(0,\bar{j})} \oplus \mathcal{H}_{\frac{4r-2+2\bar{j}}{3},(0,\bar{j})} \oplus \bar{\mathcal{H}}_{\frac{-4r+2+2\bar{j}}{3},(0,\bar{j})} \oplus (2r-2)\mathcal{L} \oplus \label{dmult} \notag \\ 
	 & \oplus \mathcal{S}_{\frac{4r+3+2\bar{j}}{3}(0,\bar{j}+\frac{1}{2})} \oplus \mathcal{H}_{\frac{4r-1+2\bar{j}}{3}(0,\bar{j}+\frac{1}{2})} \oplus \bar{\mathcal{H}}_{\frac{-4r+3+2\bar{j}}{3}(0,\bar{j}+\frac{1}{2})} \oplus (2r-2)\mathcal{L} 
\end{align}
For $\mathcal{B}$ the first line corresponds to ${\cal N} = 1$ multiplets spawning from the bottom component $|BC\rangle$, the second and the third line to the multiplets spawning from $Q |BC\rangle$ and the last one to the multiplets from $Q^2 |BC\rangle$:
\begin{align}
	\mathcal{B}_{r,\frac{R_2}{2}(0,\bar{j})}& \rightarrow \mathcal{S}_{\frac{4r+R_2}{3},(0,\bar{j})} \oplus \mathcal{H}_{\frac{4r+R_2-4}{3},(0,\bar{j})} \oplus (2r-1)\mathcal{L} \label{bmult} \notag \\
	& \oplus \mathcal{S}_{\frac{4r+R_2+1}{3},(0,\bar{j}+\frac{1}{2})} \oplus \mathcal{H}_{\frac{4r+R_2-3}{3},(0,\bar{j}+\frac{1}{2})} \oplus (2r-1)\mathcal{L}\notag \\ & \oplus \mathcal{S}_{\frac{4r+R_2+1}{3},(0,\bar{j}-\frac{1}{2})} \oplus \mathcal{H}_{\frac{4r+R_2-3}{3},(0,\bar{j}-\frac{1}{2})} \oplus (2r-1)\mathcal{L}\notag \\
	& \oplus \mathcal{S}_{\frac{4r+R_2+2}{3},(0,\bar{j})}\oplus \mathcal{H}_{\frac{4r+R_2-2}{3},(0,\bar{j})} \oplus (2r-1) \mathcal{L} 
\end{align}
Finally let us move to the semi-BPS shortened multiplets. Here for $\hat{\mathcal{C}}$ the first line corresponds to the multiplets spawning from $|BC\rangle$, the second --- to the ones spawning from $Q|BC\rangle$, the third --- from $\bar{Q}|BC\rangle$, and the fourth --- from $Q\bar{Q}|BC\rangle$:
\begin{align}
	\hat{\mathcal{C}}_{r,(j,\bar{j})} & \rightarrow \bar{\mathcal{H}}_{\frac{2}{3}(\bar{j}-j)-\frac{4r}{3},(j,\bar{j})} \oplus \mathcal{H}_{\frac{2}{3}(\bar{j}-j)+\frac{4r}{3},(j,\bar{j})} \oplus (2r-1)\mathcal{L} \oplus \label{chatmult}\notag \\
	& \oplus \bar{\mathcal{H}}_{\frac{2}{3}(\bar{j}-j)-\frac{4r+1}{3},(j+\frac{1}{2},\bar{j})} \oplus \mathcal{H}_{\frac{2}{3}(\bar{j}-j)+\frac{4r-1}{3},(j+\frac{1}{2},\bar{j})} \oplus (2r-1)\mathcal{L} \oplus \notag \\
	& \oplus \bar{\mathcal{H}}_{\frac{2}{3}(\bar{j}-j)-\frac{4r-1}{3},(j,\bar{j}+\frac{1}{2})} \oplus \mathcal{H}_{\frac{2}{3}(\bar{j}-j)+\frac{4r+1}{3},(j,\bar{j}+\frac{1}{2})} \oplus (2r-1)\mathcal{L} \oplus \notag \\
	& \oplus  \bar{\mathcal{H}}_{\frac{2}{3}(\bar{j}-j)-\frac{4r}{3},(j+\frac{1}{2},\bar{j}+\frac{1}{2})} \oplus \mathcal{H}_{\frac{2}{3}(\bar{j}-j)+\frac{4r}{3},(j+\frac{1}{2},\bar{j}+\frac{1}{2})} \oplus (2r-1)\mathcal{L};
\end{align}
For $\mathcal{C}$ the first line corresponds to $|BC\rangle$ and $Q|BC\rangle$, the second --- to $\bar{Q}|BC\rangle$, the third --- to $Q\bar{Q}|BC\rangle$, and the fourth --- to $\bar{Q}^2|BC\rangle$ and $Q\bar{Q}^2|BC\rangle$:
\begin{align}
	\mathcal{C}_{r,\frac{R_2}{2}(j,\bar{j})} & \rightarrow \mathcal{H}_{\frac{R_2+4r}{3},(j,\bar{j})} \oplus 2r\mathcal{L} \oplus \mathcal{H}_{\frac{R_2+4r-1}{3},(j+\frac{1}{2},\bar{j})}  \oplus 2r\mathcal{L} \notag \\
	& \oplus \mathcal{H}_{\frac{R_2+4r+1}{3},(j,\bar{j}+\frac{1}{2})} \oplus 2r\mathcal{L} \oplus \mathcal{H}_{\frac{R_2+4r+1}{3},(j,\bar{j}-\frac{1}{2})} \oplus 2r\mathcal{L} \notag \\
	& \oplus \mathcal{H}_{\frac{R_2+4r}{3},(j+\frac{1}{2},\bar{j}+\frac{1}{2})} \oplus 2r\mathcal{L} \oplus \mathcal{H}_{\frac{R_2+4r}{3},(j+\frac{1}{2},\bar{j}-\frac{1}{2})} \oplus 2r\mathcal{L} \notag \\
	& \oplus \mathcal{H}_{\frac{R_2+4r+2}{3},(j,\bar{j})} \oplus 2r\mathcal{L} \oplus \mathcal{H}_{\frac{R_2+4r+1}{3},(j+\frac{1}{2},\bar{j})} \oplus 2r\mathcal{L}; \qquad 
\end{align}
As $\mathcal{C}$ and $\mathcal{B}$ lie in the same equivalence classes, the decomposition rules in the equivalence class language are the same.
 
\subsubsection{Special cases}
Now let us discuss the special cases when either $r$ is small or we have some conservation at hand; we will simply list all interesting results.
\begin{itemize}
	\item[1.] $\hat{\mathcal{B}}_r$: for $r\leq 1$ the decomposition formula \eqref{bhatmult} doesn't work (it gives a negative number of long multiplets). In these cases the decomposition goes as follows:
\begin{align}
	\mathcal{B}_1 & \rightarrow \bar{\mathcal{S}}_{-\frac{4}{3}} \oplus \mathcal{S}_{\frac{4}{3}} \oplus \hat{\mathcal{H}}_{(0, 0)};  \\
	\mathcal{B}_{\frac{1}{2}} & \rightarrow \bar{\hat{\mathcal{S}}}_{(0,0)} \oplus \hat{\mathcal{S}}_{(0,0)}
\end{align}
	\item[2.] $\mathcal{D}_{r,(0,\bar{j})}$. The decomposition formula \eqref{dmult} doesn't work for $r \leq \frac{1}{2}$. $\mathcal{D}_{\frac{1}{2},(0,\bar{j})}$ contains conserved higher-spin currents, for $\bar{j}=0$ it is the multiplet that contains the conserved supercharges, and for $\bar{j}>0$ it can only appear in a free theory. The decomposition happens as follows:
\begin{equation}
	\mathcal{D}_{\frac{1}{2},(0,\bar{j})} \rightarrow \mathcal{S}_{(0, \bar{j}),4/3+\bar{j}/3} \oplus \hat{\mathcal{H}}_{(0, \bar{j})} \oplus \mathcal{S}_{(0, \bar{j}+\frac{1}{2}),5/3+\bar{j}/3} \oplus \hat{\mathcal{H}}_{(0, \bar{j}+\frac{1}{2})}.
\end{equation}
	\item[3.] $\mathcal{D}_{0,(0,\bar{j})}$: this is a generalized vector multiplet that decomposes as 
	\begin{equation}
		\mathcal{D}_{0,(0,\bar{j})} \rightarrow \hat{\mathcal{S}}_{(0,\bar{j})} \oplus \hat{\mathcal{S}}_{(0,\bar{j}+\frac{1}{2})};
	\end{equation}
	cases with $\bar{j}>0$ are possible only in a free theory.
	\item[4.] $\mathcal{B}_{r, \frac{R_2}{2}, (0, \bar{j})}$. The equation \eqref{bmult} technically stops working when $r=0$, however, in this case the restrictions are enhanced and one gets $\mathcal{E}$ instead.
	\item[5.] $\hat{\mathcal{C}}_{0,(j,\bar{j})}$. Here the usual formula \eqref{chatmult} doesn't work and we get instead
\begin{equation}
	\hat{\mathcal{C}}_{0,(j,\bar{j})} \rightarrow \hat{\mathcal{H}}_{(j,\bar{j})} \oplus \hat{\mathcal{H}}_{(j,\bar{j}+\frac{1}{2})} \oplus \hat{\mathcal{H}}_{(j+\frac{1}{2},\bar{j})} \oplus \hat{\mathcal{H}}_{(j+\frac{1}{2},\bar{j}+\frac{1}{2})}.
\end{equation}
	For $j=\bar{j}=0$ we get the stress-tensor multiplet, otherwise it is a free theory multiplet.
\end{itemize}

Looking at these results, one can notice that for an ${\cal N} = 1$ theory to be enhanced to ${\cal N} = 2$, the ${\cal N} = 1$ BPS multiplets should come in some sets. Therefore, there should be some restrictions on the ${\cal N} = 1$ index that would determine whether a theory possesses higher supersymmetry.

\subsubsection{${\cal N} = 1$ index} \label{n1index}
For reference, we list here the corrected index contributions for different ${\cal N} = 1$ multiplets; they differ from the usual ones by a factor of $(1-t/x)(1-tx)$. Our normalization is in agreement with the one in \cite{Dolan}, with the exception of considering the left index instead of right and a different definition of $SU(2)$ character $$\chi_j(x) = \frac{(x^{2j+1}-x^{-2j-1})}{(x-1/x)}$$.
\paragraph*{Free field multiplets.}
\begin{align}
	\hat{\mathcal{S}}_{(0,0)} & \rightarrow t^{2/3} \\ 
	\hat{\mathcal{S}}_{(0,\frac12)} & \rightarrow -t\left(x+\frac1x\right) + t^{2} \\ 
	\bar{\hat{\mathcal{S}}}_{(0,0)} & \rightarrow -t^{4/3} \\ 
	\bar{\hat{\mathcal{S}}}_{(\frac12,0)} & \rightarrow t^{2} 
\end{align}
\paragraph*{Conserved current multiplets.}
\begin{align}
	\hat{\mathcal{H}}_{(j,\bar{j})} & \rightarrow -(-1)^{2j+2\bar{j}}t^{2+(\bar{j}+2j)/3}\chi_{\bar{j}}(x)\\
	\hat{\mathcal{H}}_{(0,0)} & \rightarrow -t^{2}\\
	\hat{\mathcal{H}}_{(0,\frac12)} & \rightarrow t^{7/3}\left(x+\frac1x\right)\\
	\hat{\mathcal{H}}_{(\frac12,0)} & \rightarrow t^{8/3}\\
	\hat{\mathcal{H}}_{(\frac12,\frac12)} & \rightarrow -t^{3}\left(x+\frac1x\right)
\end{align}
\paragraph*{Other multiplets.}
\begin{align}
	\mathcal{S}_{R,(0, \bar{j})} & \rightarrow (-1)^{2\bar{j}} t^{R}\chi_{\bar{j}}(x)\\
	\bar{\mathcal{S}}_{R,(j, 0)} & \rightarrow 0 \\
	\mathcal{H}_{R,(j, \bar{j})} & \rightarrow -(-1)^{2\bar{j}+2j} t^{R+2j+2}\chi_{\bar{j}}(x)\\
	\bar{\mathcal{H}}_{R,(j, \bar{j})} & \rightarrow 0
\end{align}

\subsection{Relation between ${\cal N} = 1$ and ${\cal N} > 1$ equivalence classes} \label{alg}
Let us describe the general algorithm that establishes the relation between the equivalence classes. First of all, let us consider a general ${\cal N} = 1$ equivalence class (a detailed discussion can be found in appendix \ref{algN2}). It corresponds to a set of ${\cal N} = 1$ multiplets that contribute to the index in the same way and is described by $\tilde{R}_1$, $\bar{j}$ and a plus or minus sign. This means one can classify the terms in the index according to the value of $\tilde{R}_1$. For an ${\cal N} = 2$ multiplet entry $\tilde{R}_1$ can be determined as $\tilde{R}_1 = \frac{4r+R_2}{3}+2j$, for ${\cal N} = 3$ $\tilde{R}_1 = \frac{8a+ 4b + R_3}{9} + 2j$, and for ${\cal N} = 4$ $\tilde{R}_1 = \frac{3k+2p+q}{3} + 2j$. Here $r,a,b,k,p,q$ are weights in some representation (so they might be negative), and the formulas are obtained from the multiplets decomposition rules.

A general ${\cal N} = k$ equivalence class, on the other side, corresponds to a set of ${\cal N} = k$ multiplets and is described by $p_1 \ldots p_{k-1}$ that are related to the $SU(k)$ weights of the bottom component of the multiplet, $\bar{j}$, a plus or minus sign, and (not always) the charge under the $U(1)_R$ $R_k$. Under SUSY reduction an ${\cal N} = k$ multiplet $\mathcal{M}$ breaks into many ${\cal N} = 1$ multiplets (and each $\mathcal{M}$ entry breaks into many ${\cal N} = 1$ multiplet entries) that can spawn either from the bottom component of $\mathcal{M}$ or from the states that are created by the action of $Q_2, \ldots, Q_k, \bar{Q}_2, \ldots, \bar{Q}_k$ on the bottom component. As the ${\cal N} = 1$ index counts only the short multiplets that are canceled under the action of $Q_1$, only terms with maximal (or maximal and next to maximal, if we consider a multiplet with BPS shortening) $\tilde{R_1}$ in the given ${\cal N} = k$ multiplet entry may contribute.

Now let us notice, that after ${\cal N} = k \rightarrow {\cal N} = 1$ SUSY reduction the non-residual supercharges $Q_2, \ldots, Q_k$ will have charge $-\frac{1}{3}$ under $R_1$ and $\bar{Q}_2, \ldots, \bar{Q}_k$ will have charge $\frac{1}{3}$. As $Q_i$ have spin $j=\frac{1}{2}, \bar{j}=0$ and $\bar{Q}_i$ have $j=0, \bar{j}=\frac{1}{2}$, under $\tilde{R}_1=R_1+2j$ $Q_2, \ldots, Q_k$ will have charge $\frac{2}{3}$ and $\bar{Q}_2, \ldots, \bar{Q}_k$ will have charge $\frac{1}{3}$. One can also notice that ${\cal N} = 1$ multiplets that are spawned not from the bottom component of $\mathcal{M}$ can be obtained by the action of non-residual supercharges on the bottom component. This means that the ${\cal N} = 1$ multiplets that spawn not from the bottom component contribute to the index terms with bigger $\tilde{R}_1$. Therefore one can relate the ${\cal N} = 1$ equivalence class with certain $\tilde{R}_1$, $\bar{j}$ to a set of ${\cal N} = k$ multiplets, whose bottom components have the same $\tilde{R}_1$, $\bar{j}$. 

Finally, let us discuss the content of $SU(k)$ representations. A representation is described by the highest weight, the other weights can be obtained by subtracting the simple roots from it according to the well-known rules. One can notice that according to our definition of $\tilde{R}_1$, if we subtract the root that corresponds to the first line of the Cartan matrix ($2$ for ${\cal N} = 2$, $(2, -1)$ for ${\cal N} = 3$, $(2, -1, 0)$ for ${\cal N} = 4$), $\tilde{R}_1$ will decrease; subtracting other roots won't change $\tilde{R}_1$. This means that for each $SU(2)$ representation (in an ${\cal N} = 2$ multiplet) there always will be one term with maximal $\tilde{R}_1$ under ${\cal N} = 2 \rightarrow {\cal N} = 1$ reduction. For $SU(3)$ $[a; b]$ representation there will be $(b+1)$ terms, and for $SU(4)$ $[k; p; q]$ there will be $(p+1)(q+1)(1+\frac{p+q}{2})$ terms. An important implication for us is that an ${\cal N} = 2$ multiplet with bottom component $r_{R_2, (j, \bar{j})}$ will contribute 1 term to the ${\cal N} = 1$ index with $\tilde{R}_1 = \frac{4r+R_2}{3}+2j$ (and some more with larger $\tilde{R}_1$). An ${\cal N} = 3$ multiplet with bottom component $[a; b]_{R_3, (j, \bar{j})}$ will contribute $b+1$ terms to the ${\cal N} = 1$ index with $\tilde{R}_1 = \frac{8a+4b+R_3}{9}+2j$. An ${\cal N} = 4$ multiplet with bottom component $[k; p; q]_{(j, \bar{j})}$ will contribute $(p+1)(q+1)(1+\frac{p+q}{2})$ terms to the ${\cal N} = 1$ index with $\tilde{R}_1 = \frac{3k+2p+q}{3}+2j$. Using this knowledge, one can prove that with the exception of ${\cal N} = 1$ protected multiplets, it is always possible to express the contribution of a certain ${\cal N} = 1$ equivalence class to the ${\cal N} = 1$ index as a sum of contributions of an (infinite) set of ${\cal N} = 2$, ${\cal N} = 3$ multiplets and the only conditions that can be imposed on the ${\cal N} = 1$ index can be related to the ${\cal N} = 1$ protected multiplets.

The statement above contains an important loophole. In the infinite set of higher-SUSY multiplets that are used to express the contribution of an ${\cal N} = 1$ multiplet to the index some multiplets may not appear due to prohibitions on the theory content. These prohibitions are related to the conserved currents: we cannot have any spin $>2$ conserved current in a non-free theory; we can have only one stress-tensor, and the number of conserved spin-$\frac{3}{2}$ currents is also fixed. If we somehow know the flavor symmetry, we also can constrain the relevant term in the index. This idea is used to obtain a sufficient condition on SUSY enhancement from the ${\cal N} = 1$ index, as well as a necessary and sufficient condition on SUSY enhancement from the ${\cal N} = 2, 3$ index.

${\cal N} = 4$ theories do not have a $U(1)_R$ symmetry (which immediately requires the ${\cal N} = 1$ index of an ${\cal N} = 4$ theory to have only terms with $t^{k/3}$, where $k$ is an integer). Moreover, the lowest in $t$ contribution of a multiplet with a bottom component $SU(4)$ representation $[k; 0; 0]$ to the ${\cal N} = 1$ index will always have integer $\tilde{R}_1$, and the lowest in $t$ contributions of a multiplet with a bottom component $SU(4)$ representation $[k; p; q]$ for $p+q>0$ have the absolute value of the coefficient at least 3. Therefore, one can expect many more necessary conditions on an ${\cal N} = 1$ index for it to correspond to an ${\cal N} = 4$ theory.

\subsection{${\cal N} = 2$ to ${\cal N} = 1$ equivalence classes reduction relation} \label{algN2}
There are six kinds of ${\cal N} = 2$ multiplets that are canceled by the $Q_1$ supercharge (not counting the special case of free field multiplets); in Dolan and Osborn notation \cite{Osborn} they can be denoted as $\mathcal{E}$, $\mathcal{B}$, $\hat{\mathcal{B}}$, $\mathcal{D}$, $\hat{\mathcal{C}}$ and $\mathcal{C}$. Index equivalence classes were first noticed by Gadde et al. \cite{Gadde} for ${\cal N} =2$ case, and were suggested by Beem et al. in \cite{Beem} for ${\cal N} = 1$ equivalence classes. We will adopt the notation of \cite{Gadde} with slight changes.

According to Gadde et al., a general ${\cal N} = 2$ left index equivalence class, when there are no shortening conditions on $\bar{Q}$, can be described by three parameters (related to fugacities) $\tilde{r}= r + j$, $\tilde{R}_2 = R_2 + 2j$, $\bar{j}$, and a $\pm$ sign (corresponding to a positive or negative contribution to the index): 
\begin{align}
	[\tilde{R}_2, \tilde{r}, \bar{j}]_- & = \{\mathcal{C}_{\tilde{r} - j, \frac{\tilde{R}_2}{2} - j, (j, \bar{j})}, j=-\frac{1}{2},\frac{1}{2},\ldots,m \leq \tilde{r}\} \\
	[\tilde{R}_2, \tilde{r}, \bar{j}]_+ & = \{\mathcal{C}_{\tilde{r} - j, \frac{\tilde{R}_2}{2} - j, (j, \bar{j})}, j=0,1,\ldots,m \leq \tilde{r}\}
\end{align}
Here for $j=-\frac12$ $\mathcal{C}_{r,\frac{R_2}{2},(-\frac{1}{2},\bar{j})}$ denotes $\mathcal{B}_{r+\frac{1}{2},\frac{R_2+1}{2},(0,\bar{j})}$. 
In the case of symmetric shortening conditions, $R_2$ is fixed and one should define a new parameter $\hat{r} = r + j +\bar{j}$. Then the equivalence classes are:
\begin{align}
	[\hat{r}, \bar{j}]_- & = \{\hat{\mathcal{C}}_{\hat{r} - j,  (j, \bar{j})}, j=-\frac{1}{2},\frac{1}{2},\ldots,m \leq \hat{r}\} \\
	[\hat{r}, \bar{j}]_+ & = \{\hat{\mathcal{C}}_{\hat{r} - j, (j, \bar{j})}, j=0,1,\ldots,m \leq \hat{r}\}, 
\end{align}
where $\hat{\mathcal{C}}_{r,(-\frac{1}{2},\bar{j})}$ denotes $\mathcal{D}_{r+\frac{1}{2},(0,\bar{j})}$
Finally, for $\mathcal{E}$ multiplets that weren't discussed in \cite{Gadde}, we can simply define the equivalence class (every $\mathcal{E}$ lies in a class of its own) as $[e, \frac{R_2}{2}, \bar{j}]_\pm$.

For ${\cal N} = 1$ the situation is simpler: in the long multiplet decomposition into shorts, exactly two multiplets will contribute to the left index. The equivalence classes are described by two parameters $\tilde{R} = R + 2j$, $\bar{j}$, and a sign $\pm$:
\begin{align}
	[\tilde{R}, \bar{j}]_- &= \{\mathcal{H}_{\tilde{R}-2j, (j, \bar{j})}, j=-\frac{1}{2},\frac{1}{2},\ldots,m \leq \frac{3\tilde{R}}{4}-\frac{\bar{j}}{2}\}\\
	[\tilde{R}, \bar{j}]_+ &= \{\mathcal{H}_{\tilde{R}-2j, (j, \bar{j})}, j=0,1,\ldots,m \leq \frac{3\tilde{R}}{4}-\frac{\bar{j}}{2}\}	
\end{align}
Here $\mathcal{H}_{R,(-\frac{1}{2},\bar{j})}$ denotes $\mathcal{S}_{R+1,(0,\bar{j})}$.
This description leaves out $\hat{\bar{\mathcal{S}}}$, but as each of these multiplets lies in its own equivalence class and is a free field multiplet, we can disregard them for the time being. The equivalence class contribution to the index is given by 
\begin{equation}
	\tilde{\mathcal{I}}_{[\tilde{R}, \bar{j}]_\pm} = \pm (-1)^{2\bar{j}+1} t^{2+\tilde{R}}  \frac{\chi_{2\bar{j}+1} (x)}{(1-tx)(1-t/x)}
\end{equation}

To do the decomposition, one should use our results from appendix \ref{multred} and the multiplets tables from \cite{Osborn} for ${\cal N} = 2$; one should also remember that the ${\cal N} = 1$ R-charge is expressed through the ${\cal N} = 2$ R-charge as 
\begin{equation}
	R_1 = \frac{R_2}{3} + \frac{4}{3}I_3
\end{equation}

The scaling dimension restriction on the index can be reformulated as $\tilde{R}_2 \geq 2\bar{j}$ or $\tilde{R}_2 = -2$ (this case corresponds to BPS shortening on $\bar{Q}$, it can also be written as $\bar{j}=-\frac{1}{2}$ and is considered independently in \cite{Gadde}); equality corresponds to the cancellation of the multiplet's bottom component by one of the $\bar{Q}$. For ${\cal N} = 1$ scaling dimensions restrictions require $\tilde{R} \geq \frac{2(\bar{j}+2j)}{3}$ or  $\tilde{R} = \frac{4j-2}{3}$; the latter scenario corresponds to the free field multiplet. From now on we disregard the contribution of free field multiplets to the index, as they require special attention and footnotes in the analysis, while it is easy to subtract them from the index manually.

The reduction rules are slightly different for five different cases and can be formulated as
\begin{itemize}
	\item[{\bf 1.}] $\tilde{R}_2 > 2\bar{j}$. In this case the shortening conditions restrict only $Q$ and not $\bar{Q}$, the reduction goes on as
\begin{align}
	[\tilde{R}_2,\tilde{r},\bar{j}]_{(-1)^{2j}} & \rightarrow \left[\frac{4\tilde{r}+\tilde{R_2}}{3},\bar{j}\right]_{(-1)^{2j}} \oplus \left[\frac{4\tilde{r}+\tilde{R_2}+1}{3},\bar{j}\pm\frac{1}{2}\right]_{(-1)^{2j}} \oplus \notag \\ 
	&\oplus \left[\frac{4\tilde{r}+\tilde{R_2}+3}{3},\bar{j}\pm\frac{1}{2}\right]_{(-1)^{2j+1}} \oplus \left[\frac{4\tilde{r}+\tilde{R_2}+4}{3},\bar{j}\right]_{(-1)^{2j+1}}
\end{align}
When $\bar{j}=0$, the equivalence classes with $\bar{j}=-\frac{1}{2}$ that appear in the reduction should be disregarded.
	\item[{\bf 2.}] $\tilde{R}_2 = 2\bar{j}$. In this case there is a semi-BPS shortening condition on $\bar{Q}$ and the decomposition works out as (here and below $\hat{r} = r+j+\bar{j}$):
\begin{align}
	[\hat{r},\bar{j}]_{(-1)^{2j}} & \rightarrow \left[\frac{4\hat{r}-2\bar{j}}{3},\bar{j}\right]_{(-1)^{2j}} \oplus \left[\frac{4\hat{r}-2\bar{j}+2}{3},\bar{j}\right]_{(-1)^{2j+1}} \oplus \\ 
	& \oplus \left[\frac{4\hat{r}-2\bar{j}+1}{3},\bar{j}+\frac{1}{2}\right]_{(-1)^{2j}} \oplus \left[\frac{4\hat{r}-2\bar{j}+3}{3},\bar{j}+\frac{1}{2}\right]_{(-1)^{2j+1}}
\end{align}
	\item[{\bf 3.}] $\tilde{R}_2 =-2$. In this case there is a BPS shortening condition on $\bar{Q}$ and the decomposition works out as 
\begin{align}
	\left[\hat{r}, -\frac{1}{2}\right]_{(-1)^{2j}} \rightarrow \left[\frac{4\hat{r}-2}{3},0\right]_{(-1)^{2j}} \oplus \left[\frac{4\hat{r}}{3},0\right]_{(-1)^{2j+1}}
\end{align}
	\item[{\bf 4.}] $\mathcal{E}$.  In this case the decomposition works out as 
\begin{align}
	\left[e, \frac{R_2}{2}, \bar{j}\right] \rightarrow \left[\frac{R_2-6}{3},\bar{j}\right]_{-} \oplus \left[\frac{R_2-5}{3},\bar{j}\pm\frac{1}{2}\right]_{-} \oplus \left[\frac{R_2-4}{3},\bar{j}\right]_{-}
\end{align}
	\item[{\bf 5.}] Gauge multiplets $\mathcal{D}_{0,(0,0)}, \bar{\mathcal{D}}_{0,(0,0)}$ (the higher-spin multiplets can appear only in free theory). For these it is simpler to list the contributions to the ${\cal N} = 1$ index: $\mathcal{D}$ contributes $-t^{\frac{4}{3}}+t^2$, while $\bar{\mathcal{D}}$ contributes $t^\frac{2}{3}-t\chi_\frac{1}{2}(x)+t^2$.
\end{itemize}

As we have argued in appendix \ref{alg}, almost any ${\cal N} = 1$ class can be expressed as an infinite sum of ${\cal N} = 2$ classes. Let us illustrate this claim. Consider a term $\alpha t^{\tilde{R}_1+2} \chi_{\bar{j}}(x)$ in the ${\cal N} = 1$ index with $\tilde{R}> \frac{2\bar{j}}{3}$. It corresponds to the contributions of $[\tilde{R}, \bar{j}]_\pm$ classes, depending on the sign of $\alpha$, and can be expressed as the contribution of an infinite number of ${\cal N} = 2$ multiplets which are picked as follows:

If $\alpha>0$, then we first pick $\alpha$ $\mathcal{C}_{0, \frac{3\tilde{R}}{2}, (0, \bar{j})}$. Their contribution to the index is 
\begin{equation}
	\alpha t^{\tilde{R}_1+2} \chi_{\bar{j}}(x) + \alpha t^{\tilde{R}_1+\frac{7}{3}} (\chi_{\bar{j}-1/2}(x) + \chi_{\bar{j}+1/2}(x)) + O(t^{\tilde{R}_1+\frac{8}{3}}),
\end{equation}
 so we will add $\alpha$ $\mathcal{E}_{\frac{3\tilde{R}+7}{2},(0, \bar{j}+\frac{1}{2})}$ and $\alpha$ $\mathcal{E}_{\frac{3\tilde{R}+7}{2},(0, \bar{j}-\frac{1}{2})}$ to cancel the contribution from 
\begin{equation}
	\alpha t^{\tilde{R}_1+\frac{7}{3}} (\chi_{\bar{j}-1/2}(x) + \chi_{\bar{j}+1/2}(x)).
\end{equation}
This combination of multiplets will contribute to the ${\cal N} = 1$ index 
\begin{equation}
	\alpha t^{\tilde{R}_1+2} \chi_{\bar{j}}(x) + O(t^{\tilde{R}_1+\frac{8}{3}}),
\end{equation} 
then we will cancel the contributions of terms with $t^{\tilde{R}_1+\frac{8}{3}}$ by adding an appropriate number of $\mathcal{E}$ or $\mathcal{C}_0$ multiplets, and so on. For $\alpha<0$ the treatment is similar except that we start with $\mathcal{E}_{\frac{3(\tilde{R}+2)}{2}, (0, \bar{j})}$.

For $\tilde{R}= \frac{2\bar{j}-4}{3}$ only $[\tilde{R}, \bar{j}]_-$ is non-empty and it corresponds to contributions from free fields; the requirements that the free fields must obey for ${\cal N} = 1 \rightarrow {\cal N} = 2$ enhancement to happen are well-known. For $\frac{2\bar{j}}{3}>\tilde{R}> \frac{2\bar{j}-4}{3}$ only $[\tilde{R}, \bar{j}]_-$ is non-empty and for $\tilde{R} < \frac{2\bar{j}-2}{3}$ the only ${\cal N} = 2$ multiplet it can get contributions from is $\mathcal{E}$; this gives us some necessary conditions on the ${\cal N} = 1$ index for ${\cal N} = 1 \rightarrow {\cal N} = 2$ enhancement to happen.

Finally, the only set of ${\cal N} = 1$ equivalence classes we didn't take care of is $[\frac{2\bar{j}}{3}, \bar{j}]_+$ that contains $\hat{\mathcal{H}}_{(0,\bar{j})}$ multiplets. If $\bar{j}>\frac{1}{2}$, then these multiplets signify a free theory; this case is not very interesting, but we should keep in mind that if the index contains a term $t^{2+\frac{2\bar{j}}{3}} \chi_{\bar{j}}(x)$ with a plus sign, we have a free theory on our hands. If $\bar{j}=\frac{1}{2}$, then we have a supercurrent multiplet $\hat{\mathcal{H}}_{(0,\frac{1}{2})}$. This multiplet can appear only in a free theory or in a theory with enhanced supersymmetry; it can belong to an ${\cal N} = 2$ $\mathcal{D}_{\frac{1}{2},(0,0)}$ multiplet (corresponding to ${\cal N} = 2$ enhancement to higher SUSY), $\hat{\mathcal{C}}_{0,(0,0)}$ (${\cal N} = 2$ stress-tensor multiplet), $\mathcal{D}_{\frac{1}{2},(0,\frac{1}{2})}$, or $\hat{\mathcal{C}}_{0,(0,\frac{1}{2})}$ (corresponding to a free theory); appearance of this multiplet (signified by a term $t^{\frac{7}{3}}\left(x+\frac1x\right)$ with a positive overall sign) is a very strong constraint on a theory. It lies in $[\frac{1}{3},\frac{1}{2}]_+$ equivalence class, and the $[\frac{1}{3},\frac{1}{2}]_-$ equivalence class contains a $\mathcal{S}_{\frac{7}{3},(0,\frac{1}{2})}$ multiplet; if one can somehow find the number of $\mathcal{S}_{\frac{7}{3},(0,\frac{1}{2})}$ multiplets in the theory, one would be able to provide a necessary and sufficient bound on a theory either being free, being inconsistent or possessing enhanced SUSY. Finally, $\hat{\mathcal{H}}_{(0,0)}$ is a multiplet that contains a conserved spin-1 current, it can belong to ${\cal N} = 2$ $\hat{\mathcal{B}}_1$ (${\cal N} = 2$ flavor current multiplet), $\mathcal{D}_{\frac{1}{2},(0,0)}$, $\bar{\mathcal{D}}_{\frac{1}{2},(0,0)}$ (corresponding to ${\cal N} = 2$ enhancement to higher SUSY), or ${\cal N} = 2$ stress-tensor multiplet. It belongs to the $[0, 0]_+$ equivalence class, while the $[0, 0]_-$ equivalence class contains $\mathcal{S}_2$ multiplets. The index coefficient in front of $t^2$ counts the number of ${\cal N} = 1$ conserved currents minus the number of ${\cal N} = 1$ marginal deformations, this is related to a result by Green et al \cite{Green}. 

\subsection{${\cal N} = 2$ to ${\cal N} = 1$ equivalence classes reduction with flavor} \label{flavN2}
When an ${\cal N} = 2$ theory is written as an ${\cal N} = 1$ theory, its R-symmetry contains an ${\cal N} = 1$ flavor symmetry $F$. This symmetry may or may not be present in the ${\cal N} = 1$ theory we start from, but when it is, it gives us additional constraints on SUSY enhancement. We will write a general index term as $$t^{2+\tilde{R}_1} y^F \chi_{\bar{j}} (x),$$ where $F$ is the fugacity corresponding to the flavor.

\paragraph*{Sufficient conditions.}
\begin{itemize}
	\item[1. ] If the index contains a term $$\alpha t^{\frac{5}{3}}y^{-\frac{1}{3}}\chi_\frac{1}{2} (x),$$ ($\alpha$ being the overall coefficient), then there either is $N = 1 \rightarrow N = 2 + \alpha$ enhancement, or flavor is not lifted to the R-symmetry. This is because of all ${\cal N} = 2$ multiplets, only $\mathcal{D}_{\frac{1}{2}, (0,0)}$ can contribute to this term. 
	\item[2. ] If the coefficient in front of $$t^\frac{7}{3}(x+\frac{1}{x})y^{-\frac{2}{3}}$$ is positive, we have ${\cal N} = 1 \rightarrow {\cal N} = 2$ enhancement; this is because this term with a positive sign corresponds to a conserved spin-$\frac{3}{2}$ current (that comes from the ${\cal N} = 2$ stress-tensor multiplet). If we have the coefficient less than $-1$, this means that the theory has more than one stress-tensor (it has several sectors that do not talk to each other).
\end{itemize}
\paragraph*{Necessary conditions.}
\begin{itemize}
	\item[1. ] If the index contains terms with $F > \tilde{R}_1 + 2$ or $F \leq -2 -\frac{\tilde{R}_1}{2}$, SUSY enhancement cannot happen. This is because in this scenario there are no ${\cal N} = 2$ multiplets that could give such a contribution to the index; their existence is prohibited by the restrictions the ${\cal N} = 2$ superconformal invariance places on the scaling dimension. 
	\item[2. ] If the index contains terms that have non-integer $\tilde{R}_1 - F$, the SUSY enhancement cannot happen. This is because in this scenario the ${\cal N} = 2$ multiplet should have an entry with $2\cdot I_3$ non-integer, which does not make sense.
	\item[3. ] If the coefficient in front of a term $$(-1)^{2j+1} t^{2+\frac{2j}{3}} y^{1+\frac{2j}{3}} \chi_j (x)$$ for $j>\frac{1}{2}$ is positive, we have a free theory; the converse is not necessary true. This is because the only ${\cal N} = 1$ multiplet that can contribute to this term with such a sign is $\hat{\mathcal{H}}_{(0, j)}$ that can only appear in a free theory.
\end{itemize}

\section{Index and SUSY enhancement conditions derivation in 3 dimensions} \label{d3comp}
In this appendix we present various calculations and definitions, that are used in section \ref{d3}. In subsection \ref{notapp} we set the notation, in subsection \ref{d3n2} we describe multiplets content in ${\cal N} = 2$, define ${\cal N}=2$ index and shortly describe multiplet content in ${\cal N} > 2$ algebras. In subsection \ref{n3mult} we introduce notation for ${\cal N} > 2$ multiplets. In subsection \ref{n2cond} we derive conditions on ${\cal N}=2$ index for SUSY enhancement and in subsection \ref{n3cond} we do the same for ${\cal N}=3$ index. Throughout this section a significant part of the analysis is adopted from \cite{Cordova}.
\subsection{Superconformal symmetry in 3 dimensions} \label{notapp}
First of all, let us briefly set the notation. 3-dimensional superconformal algebra with $2{\cal N}$ supercharges has R-symmetry $SO({\cal N})$, and a conformal group $SO(3,2)$. Commutation relations for the supercharges are given by 
\begin{equation}
	\{Q_{r\alpha}, S^\beta_s \} = 2i(M_\alpha^\beta \delta_{rs} -i\delta_\alpha^\beta R_{rs} + \delta_\alpha^\beta \delta_{rs} D).
\end{equation}
Here we define $M^\beta_\alpha$ as 
\begin{equation}
	[M^\beta_\alpha] = \left( \begin{array}{cc}
	J_3 & J_+ \\
	J_- & -J_3
	\end{array} \right)
\end{equation}
Superconformal multiplet can be defined by its bottom component and the set of shortening conditions it obeys; bottom component is described by the charges under R-symmetry $[r_i]$, its spin $j$, and its scaling dimension $\Delta$. Shortening conditions can be divided into BPS (which can be written as $Q|BC\rangle = 0$ and require $j=0$) and semi-BPS (which prohibit one state with spin $j-\frac12$ at the second level of the multiplet).
For ${\cal N}>8$ corresponding interacting superconformal field theories cannot exist; for ${\cal N}=1$ R-symmetry group is trivial and so is the analysis (we discuss this case in section \ref{3dn1}); therefore we are restricted to $2\leq {\cal N} \leq 8$ cases. ${\cal N}=2$ case is very different from the rest, because in this case $SO(2)_R = U(1)_R$ and we have independent $Q$, $\bar{Q}$, so we will use different notation for it.

\subsection{${\cal N}=2$ multiplets}\label{d3n2}
In this case we can have shortening conditions on $Q$, $\bar{Q}$ independently and R-charge $R_2$ can be negative. The multiplet content is quite similar to $d=4, {\cal N}=1$ superconformal theories and is given by:
\begin{itemize}
	\item[1. ]{\bf BPS-semiBPS shortening}. There are only two such multiplets $\hat{\mathcal{S}}$, $\bar{\hat{\mathcal{S}}}$ with $j=0, R_2 =\pm \frac12, \Delta=\frac12$ that are the free chiral multiplets; they are absolutely protected.
	\item[2. ]{\bf semiBPS-semiBPS shortening}. The bottom component of these multiplets has $\Delta=j+1, R_2=0$, these multiplets (which we denote by $\hat{\mathcal{H}}_j$) host the conserved currents. $\hat{\mathcal{H}}_0$ hosts conserved flavor currents,  $\hat{\mathcal{H}}_{1/2}$ hosts supercurrent and $\hat{\mathcal{H}}_1$ is the stress-tensor multiplet.
	\item[3. ]{\bf BPS shortening}. The bottom component of these multiplets ($\mathcal{S}_{R_2}$, $\bar{\mathcal{S}}_{R_2}$) has $\Delta=\pm R_2, j=0$, $|R_2|>\frac12$. These multiplets host deformations and are protected for $\frac12 < |R_2| < 2$. $\mathcal{S}_1$ will be important in our studies because this multiplet appears in the decomposition of many higher-SUSY stress-tensor multiplets; we will call these multiplets BPS multiplets for short.
	\item[4. ]{\bf SemiBPS shortening}. The bottom component of these multiplets has $\Delta = 1 +j \pm R_2$, we denote these multiplets by $(\mathcal{H}_{j, R_2}$, $\bar{\mathcal{H}}_{j, R_2}$). 
	\item[5. ]{\bf Long multiplets}. These multiplets have $\Delta > 1+ j + |R_2|$,  we denote these multiplets by $\mathcal{L}_{\Delta, j, R_2}$. 
\end{itemize}
When the long multiplet hits the unitarity bound, there are several possibilities for its decomposition:
\begin{itemize}
	\item[1. ] $R_2 \neq 0$ (we consider $R_2 > 0$ for brevity, $R_2<0$ case is similar). In this case long multiplet decomposes as:
	\begin{align}
		\mathcal{L}_{j, R_2} & \rightarrow \mathcal{H}_{j, R_2} \oplus \mathcal{H}_{j-\frac12, R_2 + 1}; \quad j > 0 \\
		\mathcal{L}_{0, R_2} & \rightarrow \mathcal{H}_{0, R_2} \oplus \mathcal{S}_{0, R_2 + 2};
	\end{align}
	\item[2. ] $R_2 = 0$. In this case long multiplet decomposes as 
	\begin{align}
		\mathcal{L}_{j, 0} & \rightarrow \hat{\mathcal{H}}_j \oplus \mathcal{H}_{j-\frac12, 1} \oplus \bar{\mathcal{H}}_{j-\frac12, -1}; \quad j > 0 \\
		\mathcal{L}_{0, 0} & \rightarrow \hat{\mathcal{H}}_0 \oplus \mathcal{S}_2 \oplus \bar{\mathcal{S}}_{-2};
	\end{align}
\end{itemize}

\subsubsection{Index and equivalence classes}
One can define ${\cal N}=2$ index as
\begin{equation}
	\tilde{I}(x) = Tr ((-1)^F x^{\Delta+j})
\end{equation}
where trace is taken over all Verma module states that lie in the kernel of $\delta$; $\delta = \frac12 \{Q^\dagger, Q\}$, $Q$ is a specific supercharge, that has $U(1)_R$ charge 1 and spin $-\frac12$; $\Delta$ is the scaling dimension of the state. In practice that means that various multiplets will contribute as follows:
\begin{align}
	\tilde{I}(\hat{\mathcal{S}}) & = \frac{\sqrt{x}}{1-x^2} \\
	\tilde{I}(\bar{\hat{\mathcal{S}}}) &= \frac{-x^{3/2}}{1-x^2} \\ 
	\tilde{I}(\mathcal{S}_{R_2}) &= \frac{x^{R_2}}{1-x^2} \\
	\tilde{I}(\mathcal{H}_{j, R_2}) &= (-1)^{2j+1}\frac{x^{R_2+2j+2}}{1-x^2} \\ 
	\tilde{I}(\hat{\mathcal{H}}_j) &= (-1)^{2j+1}\frac{x^{2j+2}}{1-x^2}
\end{align}
Using these results and the decomposition rules, one can define ${\cal N}=2$ equivalence classes according to their contributions to the index:
\begin{align}
	[R_2]_+ &= \mathcal{S}_{R_2+2}, \{\mathcal{H}_{j, R_2 - 2j}, 2j=1, 3, \ldots, \lfloor R_2 \rfloor\}\\
	[R_2]_- &= \{\mathcal{H}_{j, R_2 - 2j}, 2j=0, 2, \ldots, \ldots, \lfloor R_2 \rfloor\}
\end{align}
Also $[-\frac32]_+$ contains contributions from $\hat{\mathcal{S}}$ and $[-\frac12]_-$ contains contributions from $\bar{\hat{\mathcal{S}}}$. One can notice that all multiplets contribute to the index as $$\pm\frac{x^\alpha}{1-x^2},$$ so one can introduce the notion of a corrected index $$I = (1-x^2) (\tilde{I}-1)$$ and make one-to-one correspondence between equivalence classes contents and index terms. One can also notice that $[R_2]_-$ equivalence classes are empty for $R_2<0$ (after taking into account free chirals), so index terms with $x^\alpha$ for $\alpha < 2$ give full number of $\mathcal{S}_\alpha$ multiplets.

\subsection{${\cal N} > 2$} \label{n3mult}
In this case supercharges lie in the $\mathcal{N}$ representation of $SO({\cal N})$ and shortening conditions are one-sided. The multiplet content is similar for different ${\cal N}$, we will describe it for ${\cal N}=3$. For ${\cal N}=3$ R-symmetry is $SO(3)$ and unitarity bound for long multiplet is $\Delta = j+\frac{R}{2}+1$ (R-charges of supercharges are 2,0,$-2$):
\begin{itemize}
	\item[1.] {\bf BPS shortening}. These multiplets have bottom component scaling dimension $\Delta=\frac{R}{2}$; for $R\leq 3$ they are absolutely protected. We denote them as $B^{R}$; $B^{1}$ is free hypermultiplet, while $B^{2}$ is a flavor current multiplet.
	\item[2.] {\bf semi-BPS shortening}. These multiplets have bottom component scaling dimension $\Delta=\frac{R}{2}+j+1$, we denote them as $C^{R}_{j}$.
	\item[3.] {\bf Long multiplets}. We denote these multiplets as $L^{R,\Delta}_j$
\end{itemize}
The decomposition rules are as follows: 
\begin{align}
	L^{R,R/2+j+1}_j & \rightarrow C^{R}_j \oplus C^{R+2}_{j-\frac12}; j>0 \\
	L^{R,R/2+1}_0 & \rightarrow C^{R}_0 \oplus B^{R+4}_0.
\end{align}

\subsection{${\cal N}=2$ index and SUSY enhancement} \label{n2cond}
The only way to be certain that we have SUSY enhancement is to see the presence of conserved spin-$3/2$ current that lies in $\hat{\mathcal{H}}_{\frac{1}{2}}$ multiplet\footnote{The multiplet analysis here is adopted from \cite{Cordova}.}. This multiplet can recombine with $\mathcal{H}_{-1(0)}$ and $\bar{\mathcal{H}}_{1(0)}$ into a long $d=3$ ${\cal N}=2$ multiplet $\mathcal{L}_{0,(\frac{1}{2})}^{\Delta=\frac{3}{2}}$; this means that $\hat{\mathcal{H}}_{(\frac{1}{2})}$ and $\mathcal{H}_{-1(0)}$ contribute to the index with a different sign. However, it is not the only possibility for $\mathcal{H}_{-1(0)}$ to recombine into a long multiplet: it also can recombine with $\mathcal{S}_{-3(0)}$ into $\mathcal{L}_{-1,(0)}^{2}$. This means that $\mathcal{S}_{-3(0)}$ and $\hat{\mathcal{H}}_{(\frac{1}{2})}$ contribute to the index in a very same way and we cannot extract sufficient SUSY enhancement condition from the unflavored $d=3$ ${\cal N}=2$ index.

However, there still are necessary conditions on SUSY enhancement from index. The most obvious one is that for the (corrected) index to correspond to ${\cal N}>2$ theory it must have all terms in it to be powers of $t^{1/2}$, as ${\cal N}>2$ R-symmetry groups do not have $U(1)_R$ part and cannot contribute to the index other terms. Apart from that, the necessary conditions appear due to the following reasons:
\begin{itemize}
	\item[1.] Protected multiplets and related equivalence classes. There are only three ${\cal N}=2$ protected multiplets that can contribute to the ${\cal N}>2$ index, one of which is a free field multiplet. The two remaining multiplets are $\mathcal{S}_1$ (contributes as $t$) and $\mathcal{S}_{3/2}$ (contributes as $t^{3/2}$). When the SUSY enhancement happens, $\mathcal{S}_1$ can combine with conserved current and conserved supercurrent multiplets, and some of the flavor currents may become R-symmetry currents. From that we can deduce the following conditions on the coefficients $a_1$, $a_2$ in front of $t^k$;
	\begin{align}
		{\cal N}=3:& \qquad a_1 = \mathrm{dim}\; F, \qquad a_1 + a_2 + 2 \geq 0 \\ 
		{\cal N}=4:& \qquad a_1 = \mathrm{dim}\; F, \qquad a_1 + a_2 + 5 \geq 0 \\ 
		{\cal N}=5:& \qquad a_1 = 1, \qquad a_2 \geq -9 \\
		{\cal N}=6:& \qquad a_1 = 4, \qquad a_2 \geq -14\\
		{\cal N}=8:& \qquad a_1 = 10, \qquad a_2 \geq -27
	\end{align}
	Here we list conditions on exact enhancement (e.g. ${\cal N}=3$ should be read as ${\cal N}=2$ enhanced to ${\cal N}=3$, but not ${\cal N}=4$) and also make use of the fact that there is a sufficient condition on the number of ${\cal N}=2$ conserved current multiplets (the index term $t^2$ tracks number of exactly marginal deformations minus number of conserved current multiplets) and that ${\cal N}=5-8$ theories cannot have flavor symmetry. We denote flavor symmetry group in the enhanced-SUSY theory with $F$.

	As $t^{3/2}$ contribution cannot be related to any conserved current multiplet, the SUSY enhancement conditions we can derive from it are much weaker. In fact, for enhancement to ${\cal N}=3, 4$ one cannot derive any SUSY enhancement condition for this term, while for larger SUSY they are as follows:
	\begin{align}
		{\cal N}=5:& \qquad a_{3/2} \quad \text{is divisible by 2}\\
		{\cal N}=6: & \qquad a_{3/2}  \quad \text{is divisible by 2}\\
		{\cal N}=8:& \qquad a_{3/2} \quad \text{is divisible by 4}
	\end{align}
	\item[2.] Absence of one-to-one correspondence between the ${\cal N}=2$ and ${\cal N}>2$ equivalence classes. The line of reasoning here is similar to the $d=4$ case; for ${\cal N}=3, 4$ a one-to-one correspondence can be built and no necessary conditions can be derived. For ${\cal N} = 3$ $B^{2R_2-4}$ contributes to the ${\cal N} =2$ corrected index as $x^{R_2}+O(x^{R_2+1})$, $C^{2R_2}_0$ contributes to the ${\cal N} =2$ corrected index as $-x^{R_2}+O(x^{R_2+1})$, so similarly for the $d=4$ case any term in the corrected ${\cal N} = 2$ index can be expressed as a contribution of an infinite sum of ${\cal N} = 3$ multiplets; for ${\cal N} = 4$ one can obtain any ${\cal N} = 2$ term from $B^{R,0}$ and $C^{R,0}_0$ multiplets.

	For $5\leq {\cal N}\leq 8$, however, one can get some information about the index. Let us focus on the BPS multiplets and consider a correspondence relation for them (the case for the semi-BPS multiplets can be built similarly). The $5\leq {\cal N}\leq 8$ BPS $B$ multiplets have the bottom component that is described by a rep of $SO(N)$ $[R_1, \ldots, R_k]$ and obeys the following shortening conditions:
	\begin{align}
		{\cal N}=5: &[R_1, R_2] \Rightarrow \Delta = R_1+ \frac{R_2}{2} \label{short}\\
		{\cal N}=6: & [R_1, R_2, R_3] \Rightarrow \Delta = R_1+ \frac{R_2+R_3}{2} \\
		{\cal N}=8:& [R_1, R_2, R_3, R_4] \Rightarrow \Delta = R_1+R_2 + \frac{R_3+R_4}{2} 
	\end{align}
After working out the decomposition rules one can notice that ${\cal N}=5$ $[R_1, R_2]$ multiplet will have $R_2 + 1$ ${\cal N}=2$ $\mathcal{S}_\Delta$ multiplets in its decomposition (and no other $\mathcal{S}$, $\Delta$ is defined according to the \eqref{short} for each of the ${\cal N}$), ${\cal N}=6$ $[R_1, R_2, R_3]$ will contain $(R_2+1)(R_3+1)$ ${\cal N}=2$ $\mathcal{S}_\Delta$ (and no other $\mathcal{S}$). Finally, ${\cal N}=8$ $[R_1, R_2, R_3, R_4]$ will contain $d(R_2, R_3, R_4)$ ${\cal N}=2$ $\mathcal{S}_\Delta$ and no other $\mathcal{S}$; $d(R_2,R_3,R_4)$ is the dimension of the $SO(6)$ rep with highest weight $[R_2, R_3, R_4]$. This means that for the $t^n, n\in \mathbb{N}$ contributions we can find a direct one-to-one correspondence between ${\cal N}=2$ $\mathcal{S}_n$ and higher-SUSY $B_{[n,0,\ldots,0]}$ multiplets.

The situation for the $t^{k}, k+\frac12 \in \mathbb{N}$ is more complicated, because for a state in ${\cal N}>4$ BPS (or semiBPS) multiplet to contribute in such a way to the index it must have odd $R_2$ (for ${\cal N}=5$), odd $R_2 + R_3$ (for ${\cal N}=6$) or odd $R_3+R_4$ (for ${\cal N}=8$). Therefore, using the result from the previous paragraph one can see that for ${\cal N}=5, 6$ the coefficient in front of $t^k$ must be even. For the ${\cal N}=8$ case let us consider the number of ${\cal N}=2$ BPS\footnote{The situation for semi-BPS multiplets is similar.} multiplets generated from ${\cal N}=8$ BPS:
\begin{align}
	d(R_2,R_3,R_4) = \frac{1}{12} &(R_2+1)(R_3+1)(R_4+1)(R_2+R_3+2)\cdot \notag \\
	&\cdot (R_2+R_4+2)(R_2+R_3+R_4+3)
\end{align}
As we mentioned, $R_3+R_4$ is odd, so let $R_3=2k+1$ and $R_4=2l$ without loss of generality. Then we have the following options for either even or odd $R_2$ 
\begin{align}
	d(R_2=2m) = \frac{2}{3} &(2m+1)(k+1)(2l+1)(3+2m+2k)\cdot \notag \\ 
	\cdot &(1+m+l)(m+k+l+2)\\
	d(R_2=2m+1) = \frac{2}{3} &(m+1)(k+1)(2l+1)(2+m+k)\cdot \notag \\ 
	\cdot &(3+2m+2l)(2m+2k+2l+5)
\end{align}
In the first $R_2=2m$ case, we can see that $$(k+1)(1+m+l)(m+k+l+2)$$ is even (and overall expression is divisible by 4), because for $m+l$ odd we get extra factor of 2 from $1+m+l$, for $m+l$ even and $k$ odd we get extra factor of 2 from $k+1$ and for $m+l$ even and $k$ even we get two from $(m+k+l+2)$. In the second case we can see that $$(m+1)(k+1)(2+m+k)$$ is even and overall expression is divisible by 4, because when $m+k$ is even, we get an extra factor of 2 from $(2+m+k)$ and otherwise we get it from either $m+1$ or $k+1$. As $B_{[n,0,0,1]}$ multiplet has 4 ${\cal N}=2$ $\mathcal{S}_{n+1/2}$ multiplets in its ${\cal N}=2$ decomposition, we deduce that for enhancement to ${\cal N}=8$ to be possible, ${\cal N}=2$ index coefficients $a_k$ should be divisible by 4 for non-integer k.
\end{itemize}

\subsection{${\cal N}=3$ index} \label{n3cond}

$d=3$ ${\cal N}=3$ SUSY algebra has R-symmetry group $SO(3)_R$, the charges $Q$, $\bar{Q}$ have scaling dimension $\frac{1}{2}$, spin $j=\frac{1}{2}$ and charges $2$, 0, $-2$ under the $SO(3)_R$. The multiplet content of the theory was described above in \ref{n3mult}, below we will list the ${\cal N}=3 \rightarrow {\cal N}=2$ decomposition rules for various multiplets:
\begin{align}
	B^R & \rightarrow \mathcal{S}_{R/2} \oplus \bar{\mathcal{S}}_{-R/2} \oplus \mathcal{H}_{0,R/2-1} \oplus \bar{\mathcal{H}}_{0,1-R/2} \oplus (R-3) \mathcal{L}\\
	B^2 & \rightarrow \mathcal{S}_1 \oplus \bar{\mathcal{S}}_{-1} \oplus \hat{\mathcal{H}}_0 \label{flavor} \\
	C^R_j & \rightarrow \mathcal{H}_{j,R/2} \oplus \bar{\mathcal{H}}_{j,-R/2} \oplus (R-1) \mathcal{L} \oplus \mathcal{H}_{j+1/2,R/2} \oplus \bar{\mathcal{H}}_{j+1/2,-R/2} \oplus (R-1) \mathcal{L} \\
	C^0_j & \rightarrow \hat{\mathcal{H}}_j \oplus \hat{\mathcal{H}}_{j+1/2} \label{current}
\end{align}
$B^2$ is the flavor current multiplet, $C^0_0$ is supercurrent multiplet, $C^0_\frac12$ is the stress-tensor multiplet.
${\cal N}=3$ index is written in the terms of the very same fugacities as ${\cal N}=2$, so one can simply consider contributions of ${\cal N}=3$ multiplets to the ${\cal N}=2$ index. 
The only way to be certain that we have SUSY enhancement is to see the presence of conserved spin-$3/2$ current that lies in $C^0_0$ multiplet. This multiplet can recombine with $B^4$ into a long multiplet, so there is a sufficient condition on SUSY enhancement that stems from the limitation on the number of such multiplets. From \eqref{flavor}, \eqref{current} and results in \ref{n2cond} one can see that if $-a_2 > a_1$, then there is at least ${\cal N}=3\rightarrow {\cal N}=3-a_2-a_1$ SUSY enhancement.

The necessary conditions that stem from the absence of one-to-one correspondence between the ${\cal N}=3$ and ${\cal N}>3$ equivalence classes are similar to what we had in ${\cal N}=2$ case due to the fact that all ${\cal N}=3$ multiplets contribute to the index as $\pm t^R + O(t^{R+1/2})$ and there is a one-to-one correspondence between ${\cal N}=2$ $\mathcal{S}$ multiplets and ${\cal N}=3$ $B$ multiplets as well as between ${\cal N}=2$ $\mathcal{H}$ and ${\cal N}=3$ $C$. The necessary condition on $t^{3/2}$ coefficient also stays the same.


\begin{thebibliography}{99}
\bibitem{Argyres}
  P.~C.~Argyres and M.~R.~Douglas,
  ``New phenomena in SU(3) supersymmetric gauge theory,''
  Nucl.\ Phys.\ B {\bf 448} (1995) 93
  [hep-th/9505062].
\bibitem{Romelsberger}
  C.~Romelsberger,
  ``Counting chiral primaries in N = 1, d=4 superconformal field theories,''
  Nucl.\ Phys.\ B {\bf 747} (2006) 329
  [hep-th/0510060].
\bibitem{Kinney}
  J.~Kinney, J.~M.~Maldacena, S.~Minwalla and S.~Raju,
  ``An Index for 4 dimensional super conformal theories,''
  Commun.\ Math.\ Phys.\  {\bf 275} (2007) 209
  [hep-th/0510251].
\bibitem{Bhattacharya}
  J.~Bhattacharya, S.~Bhattacharyya, S.~Minwalla and S.~Raju,
  ``Indices for Superconformal Field Theories in 3,5 and 6 Dimensions,''
  JHEP {\bf 0802} (2008) 064
  [arXiv:0801.1435 [hep-th]].
\bibitem{Dolan2}
  F.~A.~Dolan,
  ``On Superconformal Characters and Partition Functions in Three Dimensions,''
  J.\ Math.\ Phys.\  {\bf 51} (2010) 022301
  [arXiv:0811.2740 [hep-th]].
\bibitem{Romelsberger2}
  C.~Romelsberger,
  ``Calculating the Superconformal Index and Seiberg Duality,''
  arXiv:0707.3702 [hep-th].
\bibitem{Dolan}
  F.~A.~Dolan and H.~Osborn,
  ``Applications of the Superconformal Index for Protected Operators and q-Hypergeometric Identities to {\cal N} = 1 Dual Theories,''
  Nucl.\ Phys.\ B {\bf 818} (2009) 137
  [arXiv:0801.4947 [hep-th]].
\bibitem{Dolan3}
  F.~A.~Dolan,
  ``Counting BPS operators in N=4 SYM,''
  Nucl.\ Phys.\ B {\bf 790} (2008) 432
  doi:10.1016/j.nuclphysb.2007.07.026
  [arXiv:0704.1038 [hep-th]].
\bibitem{Aharony2}
  O.~Aharony, S.~S.~Razamat, N.~Seiberg and B.~Willett,
  ``3d dualities from 4d dualities,''
  JHEP {\bf 1307} (2013) 149
  [arXiv:1305.3924 [hep-th]].
\bibitem{Benini}
  F.~Benini, T.~Nishioka and M.~Yamazaki,
  ``4d Index to 3d Index and 2d TQFT,''
  Phys.\ Rev.\ D {\bf 86} (2012) 065015
  [arXiv:1109.0283 [hep-th]].
\bibitem{Ardehali}
  A.~Arabi Ardehali, J.~T.~Liu and P.~Szepietowski,
  ``Central charges from the $\mathcal{N} =$ 1 superconformal index,''
  Phys.\ Rev.\ Lett.\  {\bf 114} (2015) no.9,  091603
  [arXiv:1411.5028 [hep-th]].
\bibitem{Aharony3}
  O.~Aharony, P.~Narayan and T.~Sharma,
  ``On monopole operators in supersymmetric Chern-Simons-matter theories,''
  JHEP {\bf 1505} (2015) 117
  [arXiv:1502.00945 [hep-th]].
\bibitem{Dimofte}
  T.~Dimofte, D.~Gaiotto and S.~Gukov,
  ``3-Manifolds and 3d Indices,''
  Adv.\ Theor.\ Math.\ Phys.\  {\bf 17} (2013) no.5,  975
  [arXiv:1112.5179 [hep-th]].
\bibitem{Bashkirov2}
  D.~Bashkirov and A.~Kapustin,
  ``Dualities between N = 8 superconformal field theories in three dimensions,''
  JHEP {\bf 1105} (2011) 074
  [arXiv:1103.3548 [hep-th]].
\bibitem{Tanzini}
  S.~Benvenuti, G.~Bonelli, M.~Ronzani and A.~Tanzini,
  ``Symmetry enhancements via 5d instantons, $ q\mathcal{W} $ -algebrae and (1, 0) superconformal index,''
  JHEP {\bf 1609} (2016) 053
  [arXiv:1606.03036 [hep-th]].
\bibitem{Rastelli}
  L.~Rastelli and S.~S.~Razamat,
  ``The supersymmetric index in four dimensions,''
  J.\ Phys.\ A {\bf 50} (2017) no.44,  443013
  [arXiv:1608.02965 [hep-th]].
\bibitem{Gadde}
  A.~Gadde, E.~Pomoni and L.~Rastelli,
  ``The Veneziano Limit of N = 2 Superconformal QCD: Towards the String Dual of N = 2 SU(N(c)) SYM with N(f) = 2 N(c),''
  arXiv:0912.4918 [hep-th].
\bibitem{Beem}
  C.~Beem and A.~Gadde,
  ``The ${\cal N}=1$ superconformal index for class $S$ fixed points,''
  JHEP {\bf 1404} (2014) 036
  [arXiv:1212.1467 [hep-th]].
\bibitem{Song}
  K.~Maruyoshi and J.~Song,
  ``N = 1 Deformations and RG Flows of N = 2 SCFTs''
  arXiv:1607.04281 [hep-th].
\bibitem{Dumitrescu}
  C.~Cordova, T.~T.~Dumitrescu and K.~Intriligator,
  ``Deformations of Superconformal Theories,''
  JHEP {\bf 1611} (2016) 135
  [arXiv:1602.01217 [hep-th]].
\bibitem{Green}
  D.~Green, Z.~Komargodski, N.~Seiberg, Y.~Tachikawa and B.~Wecht,
  ``Exactly Marginal Deformations and Global Symmetries,''
  JHEP {\bf 1006} (2010) 106
  [arXiv:1005.3546 [hep-th]].
\bibitem{Parkes}
  A.~Parkes and P.~C.~West,
  ``Finiteness and Explicit Supersymmetry Breaking of the ${\cal N}=4$ Supersymmetric {Yang-Mills} Theory,''
  Nucl.\ Phys.\ B {\bf 222} (1983) 269.
\bibitem{Louis}
  J.~Louis and S.~Lüst,
  ``Supersymmetric AdS$_{7}$ backgrounds in half-maximal supergravity and marginal operators of (1, 0) SCFTs,''
  JHEP {\bf 1510} (2015) 120
  [arXiv:1506.08040 [hep-th]].
\bibitem{Intriligator}
  K.~A.~Intriligator and B.~Wecht,
  ``The Exact superconformal R symmetry maximizes a,''
  Nucl.\ Phys.\ B {\bf 667} (2003) 183
  [hep-th/0304128].
\bibitem{Agarwal}
  P.~Agarwal, K.~Maruyoshi and J.~Song,
  ``N = 1 Deformations and RG Flows of N = 2 SCFTs, Part II: Non-principal deformations'',
  arXiv:1610:05311 [hep-th].
\bibitem{Kutasov}
  D.~Kutasov, A.~Parnachev and D.~A.~Sahakyan,
  ``Central charges and U(1)(R) symmetries in {\cal N} = 1 superYang-Mills,''
  JHEP {\bf 0311} (2003) 013
  [hep-th/0308071].
\bibitem{Osborn}
  F.~A.~Dolan and H.~Osborn,
  ``On short and semi-short representations for four-dimensional superconformal symmetry,''
  Annals Phys.\  {\bf 307} (2003) 41
  [hep-th/0209056]. 
\bibitem{Papageorgakis}
  M.~Buican, T.~Nishinaka and C.~Papageorgakis,
  ``Constraints on chiral operators in $ \mathcal{N}=2 $ SCFTs,''
  JHEP {\bf 1412} (2014) 095
  [arXiv:1407.2835 [hep-th]].
\bibitem{Aharony}
  O.~Aharony and M.~Evtikhiev,
  ``On four dimensional N = 3 superconformal theories,''
  JHEP {\bf 1604} (2016) 040
  [arXiv:1512.03524 [hep-th]].
\bibitem{GMTY}
  A.~Gadde, K.~Maruyoshi, Y.~Tachikawa and W.~Yan,
  ``New {\cal N} = 1 Dualities,''
  JHEP {\bf 1306} (2013) 056
  [arXiv:1303.0836 [hep-th]].
\bibitem{Maruyoshi}
  K.~Maruyoshi and J.~Song,
  ``The Full Superconformal Index of the Argyres-Douglas Theory,''
  arXiv:1606.05632 [hep-th].
\bibitem{Sciarappa}
  P.~Agarwal, A.~Sciarappa and J.~Song,
  ``{\cal N}=1 Lagrangians for generalized Argyres-Douglas theories,''
  arXiv:1707.04751 [hep-th].
\bibitem{Benvenuti}
  S.~Benvenuti and S.~Giacomelli,
  ``Supersymmetric gauge theories with decoupled operators and chiral ring stability,''
  arXiv:1706.02225 [hep-th].
\bibitem{Benvenuti2}
  S.~Benvenuti and S.~Giacomelli,
  ``Lagrangians for generalized Argyres-Douglas theories,''
  JHEP {\bf 1710} (2017) 106
  [arXiv:1707.05113 [hep-th]].
\bibitem{Giacomelli}
  S.~Benvenuti and S.~Giacomelli,
  ``Abelianization and Sequential Confinement in $2+1$ dimensions,''
  arXiv:1706.04949 [hep-th].
\bibitem{Razamat}
  A.~Gadde, S.~S.~Razamat and B.~Willett,
  ``"Lagrangian" for a Non-Lagrangian Field Theory with $\mathcal N=2$ Supersymmetry,''
  Phys.\ Rev.\ Lett.\  {\bf 115} (2015) no.17,  171604
  [arXiv:1505.05834 [hep-th]].
\bibitem{Collingwood}
  D.~H.~Collingwood and W.~M.~McGovern,
  ``Nilpotent Orbits In Semisimple Lie Algebra: An Introduction'',
  1993,~Taylor \& Francis
\bibitem{Chacaltana}
  O.~Chacaltana, J.~Distler and A.~Trimm,
  ``Tinkertoys for the Twisted $E_6$ Theory,''
  JHEP {\bf 1504} (2015) 173
  [arXiv:1501.00357 [hep-th]].
\bibitem{Chacaltana2}
  O.~Chacaltana, J.~Distler, A.~Trimm and Y.~Zhu,
  ``Tinkertoys for the E7 Theory,''
  arXiv:1704.07890 [hep-th].
\bibitem{Shapere}
  A.~D.~Shapere and Y.~Tachikawa,
  ``Central charges of {\cal N} = 2 superconformal field theories in four dimensions,''
  JHEP {\bf 0809} (2008) 109
  [arXiv:0804.1957 [hep-th]].
\bibitem{Lie}
  M. A. A. van Leeuwen, A. M. Cohen and B. Lisser, 
  ``LiE, A Package for Lie Group Computations'', 
  Computer Algebra Nederland, Amsterdam, ISBN 90-74116-02-7, 1992
\bibitem{Bashkirov}
  D.~Bashkirov,
  ``A Note on ${\cal N}\ge 6$ Superconformal Quantum Field Theories in three dimensions,''
  arXiv:1108.4081 [hep-th].
\bibitem{Kim}
  S.~Kim and K.~Lee,
  ``Indices for 6 dimensional superconformal field theories,''
  J.\ Phys.\ A {\bf 50} (2017) no.44,  443017
  [arXiv:1608.02969 [hep-th]].
\bibitem{Cordova}
  C.~Cordova, T.~T.~Dumitrescu and K.~Intriligator,
  ``Multiplets of Superconformal Symmetry in Diverse Dimensions,''
  arXiv:1612.00809 [hep-th].
\bibitem{Bianchi}
  M.~Bianchi, F.~A.~Dolan, P.~J.~Heslop and H.~Osborn,
  ``{\cal N} = 4 superconformal characters and partition functions,''
  Nucl.\ Phys.\ B {\bf 767} (2007) 163
  [hep-th/0609179].
\end{thebibliography}
\end{document}